\documentclass[12pt,oneside,reqno]{amsart}

\usepackage{amssymb}
\usepackage{footnote}

\begin{document}


\title{Fluctuation-dissipation relations: \\ %
achievements and misunderstandings}

\author{German N. Bochkov\, and\, Yuriy E. Kuzovlev}
\address{Nijnii Novgorod %
State University, pr. Gagarin 23, 603950 Nijnii Novgorod, Russia \\  %
Donetsk Institute for %
Physics and Technology of NASU, ul. R.Luxemburg 72, %
83114 Donetsk, Ukraine}
\email{histat@rf.unn.ru , kuzovlev@fti.dn.ua}




\begin{abstract}
We discuss the ``generalized fluctuation-dissipation relations (theorems)'' %
for the first time suggested by us in 1977-1984 %
as statistical-thermodynamical consequences of time symmetry %
(reversibility) of microscopic dynamics. %
It is shown, in particular, that our old results in essence contain, %
as alternative formulations or special cases, various similar relations %
(including the ``fluctuation theorems'') what appeared %
in 1997 and later. %

\,\,\,

\end{abstract}


\maketitle

\baselineskip 18 pt

\markboth{}{}


\tableofcontents

\section{Introduction}

During last fifteen years one can observe explosively  %
growing interest in rigorous theoretical results of %
non-equilibrium statistical physics reflecting  %
fundamental properties of microscopic motion. %
One of reasons for this interest %
is discovery of new possibilities of experimental %
justification of the theory on mesoscopic level. %
Several important experiments and aspects of the underlying theory %
were reviewed not far ago in Physics-Uspekhi %
in the L.P.\,Pitaeski's article \cite{lpp}. %
However, because of its brevity, it  %
did not scope some other aspects of the subject, %
in particular, ones accompanied in the current literature %
with significant misunderstandings. %
In the present notes, we would like to highlight %
all that too.

To point out immediately what we take in mind, %
let us start from example considered in \cite{lpp}.

\subsection{On the Jarzynski and Crooks relations}

Any use of presently popular C.\,Jarzynski \cite{jar1,jar2} %
and G.\,Crooks \cite{cr1,cr2,cr3} equalities, or relations, - %
excellently expounded in \cite{lpp}, - %
presumes that a physical system under consideration %
possesses definite thermodynamically equilibrium state %
at arbitrary constant value, \,$x=\,$const\,, %
of a parameter $\,x\,$ of its Hamiltonian. %
For instance, the torsion pendulum in a liquid \cite{lpp} %
finds definite equilibrium position (with equilibrium %
fluctuations around it) at any constant value %
of the torque $\,x\,$. Then any given \,$x=\,$const\,  %
determines definite value \,$F(x)\,$\, of free energy %
of the system, as characteristics of its corresponding %
equilibrium state, and hence one can speak about %
changes of free energy,  $\Delta F$\,. For example, %
when initial, at \,$t=0$, equilibrium state of the system  %
is perturbed by some variations of its parameter, %
\,$x(t)$\,, and the Jarzynski and Crooks (J-C) equalities %
connect system's energy fluctuations in this %
process to quantity $\,\Delta F = F(x(t))-F(x(0))\,$.

\subsection{On peculiarities of open systems}

Imagine now that pendulum's hanger was made movable %
(inserted into bearing), so that the pendulum turned  %
to rotator. The resulting system, - analogue of rotary  %
viscosimeter, - can stay in equilibrium %
under zero value of the torque only, \,$\,x=0\,$. %
If $\,x\neq 0\,$, then this systems %
becomes driven to non-equilibrium %
(dissipative) state where the rotator %
constantly goes round. In such a state, %
system's free energy has no certain value, %
merely because it has no general  theoretical %
definition for thermodynamically non-equilibrium %
systems. Consequently, its changes \,$\Delta F$\, also  %
are not defined.  %

Similarly, a conducting medium can stay in equilibrium %
only under zero value of such its Hamiltonian parameter %
as electric field, and the aforesaid concerns also  %
this system. Such the systems we here will term ``open'' %
while their opposition ``closed''.

From above discussion it follows that the J-C equalities %
can be applied to open systems at exclusive time moments %
only when condition of ``cyclic process'' %
(the term from \cite{jar}, see below) is satisfied:\, %
$\,x(t)=x(0) =0\,$. This fact makes J-C equalities %
practically fruitless in respect to non-equilibrium states %
(see paragraphs 2.7 and п.2.9). %

\subsection{On old (fluctuation-dissipation) %
relations}

Exact relations applicable to open systems %
without any additional conditions  %
were obtained more than thirty years ago %
in our works \cite{bk1,bk2,bk8,bk3,bk4,bk5}. %
They are valid at every time moment %
regardless of current values of parameters, %
and therefore constantly produce information %
about a non-equilibrium state. %
Obviously, objects like free energy changes, %
$\,\Delta F\,$, in principle could not appear %
in these relations. Due to this their property, %
for their use it is sufficient if an equilibrium %
state exists at least at one point (when $\,x=0\,$). %
Therefore, they extend also to closed systems %
to which in fact are addressed J-C equalities. %
This advantage of our relations %
appears because they deal with fluctuations %
of not full energy of a system but its part %
without energy of interaction with surroundings %
(responsible for the parameters), that is  %
internal energy, which is more closely connected to %
dissipation (see p.2.1 and p.2.2).

\subsection{On misunderstandings}

The aforesaid helps to understand %
why we classify as misunderstanding  %
the opinion migrating in the literature %
(see e.g. review article \cite{rmp} and %
related references in \cite{z12}) %
and claiming that our relations are %
particular cases of J-C relations %
(or the latter ``reduce'' to our %
ones in particular cases). %

How such misunderstandings do arise,  %
one can see from Jarzynski's article %
\cite{jar} where the author compares %
old (our) and new (J-C) relations and concludes %
that  they are equivalent in special case %
of ``cyclic processes''\, %
\footnote{\, %
``For the special case of {\it cyclic} processes, in which %
the perturbation is turned on and then off, Eqs. (1) and (2) are %
equivalent'' (\cite{jar}, p.496). %
Here the equalities (\ref{ej}) and (\ref{ebk}) %
are taken in mind. Below we will compare them from our %
viewpoint in p.2.7, 2.9, and 3.3-5.}\,. %
We can agree with this statement, but  %
in purely formal sense only and only %
in respect to closed systems.  %
The matter is that in \cite{jar} it was %
assumed that systems under consideration %
possess equilibrium states at any values of %
their parameters. In other words, existence of open %
systems was not taken into account in \cite{jar}. %
Hence, the comparison was made beside the point, %
by surface signs, such as appearance of the quantity %
$\,\Delta F$ (see above) which is important component %
of new (J-C) relations but is absent in old (our) ones. %
From viewpoint of \cite{jar} this difference looks  %
as defect of our theory which gets rid of it  %
in the case of ``cyclic processes'' only,\, %
\footnote{\, %
In fact, we never attracted %
``cyclic'' or any other restrictions on %
time variations of parameters. %
One can easy verify this fact due to %
easy availability of main our works  \cite{bk1,bk2,bk8} %
through the internet.}\,, %
when in the J-C theory \,$\Delta F =0$\,. %
By these reasons, it is not surprising %
that the formally neutral conclusion of \cite{jar} %
is interpreted by readers of this paper as  %
indication of particular character %
of our results. %

More precisely, one can meet publications  %
\cite{cht,cth} free of so free interpretations,  %
but qualitative differences between the two %
types of systems are not accentuated there too. %
Thus, the subject really needs in discussion.

We do not pretend to review of generalized %
fluctuation-dissipation relations,  - %
as we think of the subject, - but hope for %
usefulness of the following notes  %
for interested readers. %
For simplicity and brevity we confine our %
consideration to classical mechanics, %
at that keeping parallels to our %
remarks from \cite{z12} and the short review \cite{lpp} %
(it strongly stimulated us, %
for which we are grateful to its author). %

\subsection{On history of the subject}

Preliminarily, we would like to recall %
that interest in rigorous results of  %
statistical mechanics has very long history. %
They include the Kirchhoff law \cite{ll5}, %
Einstein relation \cite{lp,ein}, Nyquist formula \cite{ll9,nyq} %
and the unifying fluctuation-dissipation theorem %
(FDT) \cite{ll5,cv}. %
Later, the Efremov's ``quadratic FDT'' \cite{efr}, %
Stratonovich's ``four-index relations'' \cite{str,strb} %
and his Marcovian nonlinear fluctuation-dissipation %
theory \cite{strb,strm} had appeared. %
We in 1977-1981 for the first time obtained \cite{bk1} %
and investigated \cite{bk2,bk8,bk3,bk4} %
the ``generalized fluctuation-dissipation relations'' (FDR), %
or theorems \cite{bk3}, %
in an universal way connecting probabilities %
of observation of mutually time-reversed processes %
and changes of system's entropy during these processes. %
The first of such relations was formula (7) from \cite{bk1}.  %
And most general of them is equality (2) from \cite{bkt}, %
\begin{eqnarray}
P(\Pi_+)\,\exp{[-\Delta S(\Pi_+)]}\, %
=\,P(\Pi_-) \,\,,\,\label{mfds}
\end{eqnarray}
where symbol\, $\,\Pi_+\,$ denotes some process, %
i.e. a complex of results of observations and %
measurements of definite sorts which can realize %
in the given system under given conditions  %
(concerning initial state and external perturbations %
of the system),\, $\Pi_-\,$ is time reversal of %
$\,\Pi_+\,$ (at that the reversion applies to both %
the results and conditions),\, %
$\,P(\Pi_\pm)\,$ are probabilities of realization of these %
processes (to be precise, results of the %
measurements), and\, %
$\,\Delta S(\Pi_+) =-\Delta S(\Pi_-)\,$ is %
system's entropy change in the forward process. %
This formula covers both closed and open systems, %
and extends to processes which (not only finish in %
but also) start from thermodynamically non-equilibrium %
states (see p.2.10 and p.3.2). %
In the same works and in \cite{bk5,pj,bk6,bk7,bk9,bk10} %
we and later in \cite{i1,i3,i2,chr,fdr,dvr} one of us %
considered other forms of FDR (first of all,  %
in terms of characteristic functionals) %
and various their consequences and applications. %

Notice that formula (\ref{mfds}), published in 1984 %
as a resume of our results, exceeds analogous relations, %
including ``fluctuation theorems'' \cite{rmp,cht,cth,pt}, %
published in 1997 and later by Jarzynski, Crooks and %
their followers. %
This statement does not minimize the importance of %
Jarzynski and Crooks works which introduced newly %
theoretically attractive and practically useful %
forms of FDR (see p.2.2, 2.4, 3.2, 3.3) %
and initiated the today's ``boom'' in this field. %

\section{Statistical equalities for non-equilibrium processes}

\subsection{Hamiltonians, their parameters, %
and two types of systems}

We will speak about Hamiltonian dynamical systems %
being under external influences. The latter %
are described by parameters  \,$\,x\,$\, of %
systems' Hamiltonians,\, \,$H(q,p,x)=H(\Gamma,x)$\,, %
where \,$\{q,p\}=\Gamma $\, are canonical microscopic %
variables. The parameters can vary with time  %
by a given law. The \,$\Gamma $\,'s values at arbitrary  %
time moment, \,$\Gamma(t)$\,, are in definite  %
one-to-one relationship, - determined by the Hamilton %
equations of motion \cite{arn}, - with values at any other %
time moment, for instance, with \,$\Gamma(0)\equiv \Gamma$\,. %
According to the Hamilton equations, system's full energy, %
$\,H(t)=$ $H(\Gamma(t),x(t))$\,, changes only if %
the parameters are changing, so that
\begin{eqnarray}
\frac {dH(t)}{dt}\, =\, %
\frac {dx(t)}{dt}\cdot %
\frac {\partial H(\Gamma(t),x(t))} %
{\partial x(t)}\, =\, %
- \frac {dx(t)}{dt}\cdot Q(t)\,\, \label{ddh}
\end{eqnarray}
Here \,$Q(t)= Q(\Gamma(t),x(t))$\, and %
\,$Q(\Gamma,x)= -\partial H(\Gamma,x)/ %
\partial x$\, are system's internal variables  %
conjugated with the parameters.

If a system is closed, then changes of its full energy  %
$\,H(t)$\, quite reasonably characterize changes  %
of system's state, as it is in the J-C relations. %
The openness of a system, - in the sense %
outlined in the Introduction, - %
presumes that at \,$x\neq 0$\, it may
accept from its surroundings  %
(sources of external influences)  %
an unboundedly large amount of energy, %
even if the parameters \,$x$\, are kept constant, \,$x=\,$const\,. %
At that, however, according to (\ref{ddh}), %
the full energy $\,H(t)\,$ also stays constant. %
This means that changes of one of its parts, - %
its internal, or intrinsic, energy, - %
are compensated by changes of another part, %
namely, energy of system's interaction with %
surroundings. Thus, now $\,H(t)\,$ is not adequate %
characteristics of system's state, and %
it is more meaningful to deal, instead of it, %
with system's internal energy, \,$H_0(t)\,$. %

A typical inter-connection between %
\,$H(t)\,$ and  \,$H_0(t)\,$ can be illustrated %
by the examples of torsion pendulum and rotator %
from Introduction. %
Evidently, for both them the rate of change %
of internal energy $\,H_0(t)\,$  is nothing but %
the work produced by the torque \,$x(t)$\, %
per unit time:
\begin{eqnarray}
\frac {dH_0(t)}{dt}\, =\, x(t)\, %
\frac {dQ(t)}{dt} \,\,, \, \label{ddh0}
\end{eqnarray}
where\, $Q(t) $\, means their rotation angles. %
At the same time, the rate of change of %
$\,H(t)\,$ is expressed by (\ref{ddh}) %
with \,$Q(t)$\, being the same rotation angle %
(see formulae (5) and (25) in \cite{lpp}). %
After subtracting (\ref{ddh0}) from (\ref{ddh}), %
one has for the interaction energy, \,$H-H_0$\,, %
equation $\,d[H-H_0]/dt =$ $-\, d[x\,Q]/dt\,$, %
so that \,$H(t)-H_0(t)=$ $-x(t)\,Q(t)$\, %
accurate to an arbitrary constant. %

It is  natural to supplement this with %
supposition that \,$H_0(t)=H_0(\Gamma(t))$\,, that is %
internal energy has no direct dependence %
on \,$x$\,. Then from arbitrariness of function %
\,$x(t)$\, and arbitrariness of phase trajectory $\,\Gamma(t)$\, %
it follows for both the systems that
\begin{eqnarray}
H(\Gamma,x) \, =\, H_0(\Gamma^\prime,Q) - %
x\cdot Q \,=\, %
\nonumber \\ \,=\, %
H_0(\Gamma) - %
x\cdot Q(\Gamma) \,\,, \, \label{lh}
\end{eqnarray}
with \,$Q(\Gamma)$\, also having no direct %
dependence on \,$x$\,. %
Here in the top raw the angle \,$Q$\, %
is considered as independent (canonical) variable,  %
and \,$\Gamma^\prime $\, denotes all the rest  %
of variables. In the bottom raw, it is presumed %
that generally \,$Q $\, can be treated  as a function %
of different canonical set of variables (then $Q(t)=Q(\Gamma(t))\,$). %
Hamiltonians what look like (\ref{lh}) %
can be termed ``bilinear'', since %
the interaction with surroundings there %
is linear separately in respect to the parameters %
\,$x$\, and the conjugated internal variables \,$Q$\,. %

Clearly, a difference between pendulum and rotator %
is that rotator is able to make arbitrary number of %
full turns, that is its angle \,$Q(t)\,$ %
may vary over infinite range. %
In (\ref{lh}) this difference is invisible\, %
\,\footnote{\,
Nevertheless, already in 1977 interest %
in open systems was so wide that explanations of %
their existence seemed unnecessary that time.}\,, %
since hidden in `the `eigen'' system's %
Hamiltonian \,$H_0(\Gamma^\prime,Q)$\,. %
It can either include (for pendulum) or not include %
(for rotator) an elastic contribution unboundedly %
growing as \,$Q$\, grows, for instance, \,$c\, Q^2/2\,$ %
(if elasticity of pendulum's wire or ribbon %
obeys the Hooke's law).  %
If it is absent (in case of rotator) then %
\,$H_0(\Gamma^\prime,Q\pm 2\pi)=$ $H_0(\Gamma^\prime,Q)$\,,  %
therefore arbitrary large translations of %
\,$Q$\, may change the Hamiltonian by an  %
immaterial constant only. %

It is obvious that all open systems (OS)  %
interact with surroundings through such   %
variables, indifferent to shifts, %
and therefore Hamiltonians of OS %
are naturally bilinear. %

Let us consider differences between OS %
and closed systems (CS) from the point of view of %
statistical mechanics, where one can not  %
do without (density of) probability distribution %
of microscopic states of the system, %
\,$D(q,p;t) = D(\Gamma;t)$\,. %
Following principles of the Gibbs statistical mechanics, %
we have to represent thermodynamically equilibrium  %
states of systems with constant parameters  %
by the classical canonic distributions

\begin{eqnarray}
D_{eq}(\Gamma,x)= %
\exp{\{[F(x)-H(\Gamma,x)]/T\}} \,\,,\label{eq}
\end{eqnarray}
where \,$T$\, is system's temperature (in energy units), %
and \,$F(x)$\, is the above mentioned free energy, %
to be determined from the probability normalization %
condition:\, $\int D(\Gamma;t)\, d\Gamma =1$\,. %

In case of torsion pendulum, for instance, %
with the Hooke's elasticity, %
one finds from (\ref{eq}) and (\ref{lh})  %
\,$\Delta F =$ $F(x)-F(0)$ $=-x^2/2c\,$ %
(formula (26) in \cite{lpp}).  %
In case of rotator, the Hamiltonian (\ref{lh}) becomes  %
a linear function of \,$Q\,$, and integral along \,$Q$\, %
axis in (\ref{eq}) diverges. At \,$x=0$\, the divergency is %
linear, therefore distribution (\ref{eq}) keeps a %
probability-theoretical meaning as limit of uniform %
along \,$Q$\, distribution. However, if \,$x\neq 0$\, %
then the divergency is exponential, and expression (\ref{eq}) %
no more allows a reasonable probabilistic interpretation. %
In other words, at \,$x\neq 0$\, such a system has no %
equilibrium states, and there are no grounds to speak %
about its free energy.

\subsection{Two types of parameters and %
inter-relations of ``old'' and ``new'' results}

Of course, not all application can be satisfied %
by the simple bilinear Hamiltonians like (\ref{lh}). %
First of all, if take in mind CS or ``mixed'' systems %
what are open in respect to some  of several their parameters %
but closed in respect to others. %
However, any Hamiltonian which is ``good'' enough %
function of its arguments can be written in the form
\begin{eqnarray}
H(\Gamma,x) = H_0(\Gamma) - h(\Gamma,x) \,\,, \label{fh}
\end{eqnarray}
with condition \,$h(\Gamma,0)=0\,$ ensuring unambiguity %
of this decomposition.  %
At that, the point of origin  \,$x=0$\, %
in parameters' space of CS may be defined %
in any suitable way, e.g. as point of extremum %
of free energy, where \,$\partial F(x)/\partial x =0$\,, %
thus corresponding to ``unperturbed system''.

Just for such Hamiltonians were deduced %
main results of our works \cite{bk1,bk3}, - %
as it was clearly pointed out there, - although %
for better visualization of formulae most %
of them were displayed in terms of bilinear %
Hamiltonians\, %
\footnote{\, %
Notice, however, that any ``good'' Hamiltonian %
can be represented in bilinear form (\ref{lh}) if treating \,$x$\, %
as a proper set of effective parameters (functions of  %
actual parameters) and \,$\cdot$\, as their %
contraction (``scalar product'') with corresponding set %
of phase functions. %
}\,. %
Anyway forms (\ref{lh}) or (\ref{fh}) comprise systems %
with parameters of the above mentioned type, which can be %
named ``force parameters'' (FP), or ``forces''  %
since they frequently represent forces in the sense %
of physical mechanics, or their potentials, or fields %
(or their sources if fields themselves are  %
constituent part of a system). %

Some complements can be required %
if Hamiltonian decomposition like (\ref{fh}) %
occurs superfluous, since \,$H(\Gamma,x)$\, by itself already %
represents system's internal energy, or impossible,  %
since \,$H(\Gamma,x)$\, appears too singular function %
of some of its arguments.

For example, when \,$x$\, is position of movable edge %
of a spring immersed into liquid (thermostat). %
In the experiments with ribonucleic acids (RNA), %
well described in \cite{lpp}, role of the spring %
is played by a pack of RNA molecules. %
At that,  \,$Q=-\partial H/\partial x$\, acquires %
meaning of a force acting from the spring %
onto a transmitter of external influence %
(``actuator''), so that (\ref{ddh}) is just %
the external work (per unit time) against the system. %

Another example gives wine in a wineskin %
whose disposition and deformations serve as parameters %
of this system. Singularity of Hamiltonian (and especially %
Poisson brackets) here, as well as in the previous %
example, is due to that changes of Hamiltonian %
parameters are simultaneously changes of its %
domain of definition in the phase space.

Parameters of such the type,  - which determine positions %
of some elements or boundaries of a system, - %
can be named  ``positional parameter'' (PP). %
They must change in a continuous way, %
since their instant change would mean %
infinitely fast displacements of some parts of system. %
A decomposition like (\ref{fh}) is incorrect %
in respect to such parameters, in view of that %
domains of definition of \,$H(\Gamma,x\neq 0)$\, %
and \,$H(\Gamma,0)=$ $H_0(\Gamma)$\, are different.

In opposite, the ``force parameters''  %
are not liable to be continuously varying %
and principally have all rights to make instant jumps, %
since their jumps do not change current microscopic %
state of a system, instead they merely somehow redirects %
its further evolution. %
For example, the force parameter (torque) of torsion %
pendulum can be instantly ``switched''  %
in theory and practically instantly in experiment, %
due to possibility of fast enough operation with electric current %
(in magnetic field) what creates the torque (see \cite{lpp,tm}).  %

Parameters of OS certainly belong %
to this ``force'' type. Indeed, %
if any deviation of parameter from zero %
drives a system into constant motion, %
then speed of the deviations is not %
essential for its behavior. %

From all the aforesaid it follows that %
neither first (from 1997-1999) %
new (J-C) results nor first (from 1977-1979) %
old our (B-K) results are quite %
``all-embracing'', and their relations to reality %
can be reflected by table
\begin{eqnarray}
\begin{array}{cccc}
\texttt{   } & \texttt{OS} & \texttt{CS-FP} & \texttt{CS-PP} \\ %
\texttt{J-C (1997-99)} & \texttt{-} & \checkmark & \checkmark \\ %
\texttt{B-K (1977-79)} &  \checkmark & \checkmark & \texttt{-} %
\end{array} \label{rel}
\end{eqnarray}
Here the check mark means applicability in general, %
the hyphen means applicability under special %
conditions only, and the abbreviations %
``OS'', ``CS'', ``FP'' and ``PP'' were introduced before. %

However, our slightly more late relation (\ref{mfds})  %
in full measure covers also third column of (\ref{rel}), %
as well as FDR (2.25)-(2.26) from \cite{bk3} %
which were intended for CS and are completely %
transferable to the case of PP since in fact %
do not resort to decomposition (\ref{fh}). %
On the other hand, today's followers of Jarzynski and Crooks %
move to the first column of (\ref{rel}).

\subsection{The Liouville theorem and   %
statistical equalities}

Let us go to our comparison of ``new'' and ``old'' %
relations. All that wholly or substantially %
result from the Liouville theorem \cite{arn} %
saying that Jacobian of canonical %
variables' transformation from \,$\Gamma =\Gamma(0)$\, %
to \,$\Gamma(t)$\, always equals to unit,\, %
\,$d\Gamma/d\Gamma(t)\,=1$\, %
(phase volume conserves). %

For beginning, it will help us to derive one trivial %
but significant statistical identity. %
Let \,$D_1(\Gamma)$\, and \,$D_2(\Gamma)$\, be %
some two probability distributions, both normalized %
to unit and nowhere turning to zero. %
Notice first that, because of the Liouville theorem, %
\,$\int D_2(\Gamma(t))\, d\Gamma =$  %
$\int D_2(\Gamma(t))\, d\Gamma(t)$ $=1 $\,. %
Second, dividing and multiplying the integrand here %
by \,$D_1(\Gamma)$\,, we have
\begin{eqnarray}
\int \exp{\left[\, \ln\,\frac {D_2(\Gamma(t))} %
{D_1(\Gamma)}\,\right]}\, %
D_1(\Gamma)\, d\Gamma\, =\,1 \,\, \label{tj0}
\end{eqnarray}
Third, inserting here in place of \,$D_1$\, and %
\,$D_2$\, two of canonical distributions (\ref{eq}), - %
\,$D_1(\Gamma)=D_{eq}(\Gamma,a)$\, and  %
\,$D_2(\Gamma)=D_{eq}(\Gamma,b)$\, with some  %
\,$a$\, and \,$b$\,, - we come to equality
\begin{eqnarray}
\exp{\{[F(b)-F(a)]/T\}}\, \label{tj} %
\times\, \\ \times \nonumber %
\,\langle\, %
\exp{\{-[H(\Gamma(t),b)- %
H(\Gamma,a)]/T\}}\,\rangle_a \,=\,1 %
\,\,,
\end{eqnarray}
where the angle brackets denote  %
averaging over canonical distribution of %
initial conditions \,$\Gamma=\Gamma(0)$\, %
of phase trajectory  \,$\Gamma(t)$\,:
\begin{eqnarray}
\langle \,\dots\,\rangle_x \,= %
\int \dots\, D_{eq}(\Gamma,x)\, d\Gamma\, \,, \, \nonumber
\end{eqnarray}
with the tag under brackets showing parameters %
of the initial distribution.

Identities like (\ref{tj}), or (\ref{sf}) %
from p.3.2 below, are satisfied regardless of %
magnitudes and rates of time variations of parameters %
and degree of non-equilibrium induced by them,  %
therefore, such identities indicate existence of %
universal relations (just what we call ``FDR'') %
between characteristic (average or most probable) %
direction of system's evolution and accompanying %
non-equilibrium fluctuations.

\subsection{Jarzynski equality (JE)}

Let us choose in identity (\ref{tj}) \,$a=x(0)$\, and \,$b=x(t)$\,.  %
Then it turns to the ``Jarzynski equality'' \cite{lpp,jar1}:

\begin{eqnarray}
\exp{[\Delta F(t)/T]}\,\,\langle\, %
\exp{[\,-W(t)/T\,]}\, %
\rangle_{x(0)}\, =\,1\, \,, \, \label{ej}
\end{eqnarray}
where \,\,$\Delta F(t) =F(x(t))-F(x(0))\,$\,, %
and \,$W(t)$ is change of system's full energy %
during observation time. According to (\ref{ddh}),
\begin{eqnarray}
W(t) = H(\Gamma(t),x(t)) - %
H(\Gamma,x(0))\,= \label{w} %
 \,\\ = \nonumber %
-\int_0^t  \frac {d x(t^\prime)} %
{d t^\prime} \cdot Q(t^\prime) \, dt^\prime\,\, \,
\end{eqnarray}
with \,$Q(t)=Q(\Gamma(t),x(t))$\, and %
\,$Q(\Gamma,x) =$ $-\partial H(\Gamma,x)/\partial x$ %
$=\partial h(\Gamma,x)/\partial x$\,.

The Jarzynski equality (JE) is applicable to any CS,  %
including the case of PP (see p.2.2). %
But in respect to OS it makes sense %
only if \,$x(0)=0$\, and, besides, in  %
the present time moment also \,$x(t)=0$\, %
(i.e. external influences disappear), %
since in case of OS at \,$x(0)=a\neq 0$\, %
the required in (\ref{tj}) canonical initial distribution  %
does not exist, and at \,$x(t)=b\neq 0$\, the required %
free energy \,$F(b)$\, is not defined. %
At \,$x(t)=x(0)=0$\,, when JE has a meaning, %
it coincides with our equality (\ref{ebk}).

\subsection{Bochkov-Kuzovlev equality (BKE)}

Now, let us choose in (\ref{tj}) \,$a=b=0$\,. %
In case of OS, as we already know, this is %
the only allowable choice. %
Then identity (\ref{tj}) implies
\begin{eqnarray}
\langle\,\exp{[\,-E(t)/T\,]}\, %
\rangle_0\,=\,1\,\,, \, \label{ebk}
\end{eqnarray}
where \,$E(t) = H_0(\Gamma(t)) - %
H_0(\Gamma)\,$\, is change of internal, or intrinsic, %
energy of a system during its observation. %
By the Hamilton equations of motion, %
\begin{eqnarray}
E(t)= \label{e} %
\int_0^t   \frac {d\Gamma(t^\prime)}{dt^\prime}\cdot %
\frac {\partial h(\Gamma(t^\prime),x(t^\prime))} %
{\partial \Gamma(t^\prime)} \, dt^\prime\,
\end{eqnarray}
If Hamiltonian is of the bilinear type (\ref{lh}) %
then this expression simplifies, in accordance with (\ref{ddh0}), to
\begin{eqnarray}
E(t) = \label{e0} %
\int_0^t  x(t^\prime)\cdot \frac {d Q(t^\prime)} %
{d t^\prime} \, dt^\prime\,\,
\end{eqnarray}
At that, generally speaking, \,$x(0)\neq 0$\,, %
that is external forces are not assumed to be %
absent in the beginning of observation.

Equality (\ref{ebk}) for the first time appeared %
in \cite{bk1} from more general relations which involve %
time-reversed evolution (see Section 3). Evidently, just %
it was mentioned in \cite{lpp} as ``Bochkov-Kuzovlev equality'' (BKE).

In opposite to JE, BKE is freely applicable to OS at any %
time moment, independently on current values of parameters %
\,$x(t)$\,, and to steady non-equilibrium (dissipative) %
states. What is for CS, BKE is freely applicable to them %
in case of FP, but in case of PP under special conditions %
only, \,$x(t)=x(0)=0$\, (since otherwise domain of definition %
of \,$\Gamma(t)$\, in (\ref{tj}) would be different from %
that of  \,$\Gamma(0)$\,, and the latter from domain %
of definition of \,$H_0(\Gamma)$\,). %
At that, BKE coincides with JE.

\subsection{On physical interpretation of  %
statistical equalities and jumps of external forces}

The tag ``\,$x(0)$\,'' under angle brackets in JE (\ref{ej}) %
emphasizes that parameters of initial (at \,$t=0$) %
canonical distribution of micro-states of a system coincide %
with parameters of its Hamiltonian at the initial time moment. %
Usually it goes without saying, and it is thought that %
JE presumes a system which before \,$t=0$\, was in equilibrium state %
with constant parameters equal to \,$x(0)$\,. %

But in fact parameters of initial distribution in no way %
affect behavior of one or another concrete phase trajectory, %
either before or after \,$t=0$\,. Therefore, firstly, %
the non-coincidence \,$x(0)\neq a$\, in (\ref{tj}) is possible %
and by itself does not say that at \,$t=0$\, there is jump %
of Hamiltonian parameters from \,$a$\, to \,$x(0)$\,. %
Secondly, system's equilibrium before \,$t=0$\, is %
additional independent assumption not specified automatically  %
by the tag. Without it, just same canonical %
distribution may appear in theory in the role of model %
characteristics of non-equilibrium states (see p.2.10).

On the other hand, an experimental testing of JE or BKE  %
(see p.2.8) indeed needs in practical realization of  %
mentioned additional condition, %
that is in relaxation to equilibrium, or %
``thermalization'', of system (at \,$t<0$\,) being %
governed by Hamiltonian with parameters equal to \,$a$\, %
(otherwise, it would be unreal to organize a sampling of %
experiments adequately corresponding to \,$D_{eq}(\Gamma,a)$\,). %
Then, jumps \,$ x(0) \neq a$\, acquire literal sense. %
But, as we underlined in p.2.2, this is quite %
rightful behavior of FP, since their jumps do not destroy %
continuity of time evolution of (canonical) %
microscopic variables. %

By these reasons, cases of discontinuities, or jumps, %
of the force parameters (FP), - such as \,$x(0)\neq a$\, %
in (\ref{tj}) or \,$x(0)\neq 0$\, in (\ref{ebk}), - %
are not less important for theory and its applications\, %
\footnote{\, %
Variation of jump-like (piecewise constant) %
dependencies \,$x(t)$\,, along with variational differentiation %
in respect to \,$x(t)$\,, helps to transform generalized %
FDR (statistical equalities) like (\ref{mfds}) or (\ref{ebk}) %
into various relations between linear and non-linear response %
functions (Green functions, susceptibilities, conductances, etc.) %
and irreducible second-, third- and higher-order statistical %
correlations (cumulants) of fluctuations %
(for examples see \cite{bk1,bk2,bkt,bk7,bk9,i1,i3}). %
}\,, %
than the case \,$x(0)=a$\, assumed in JE. %
In this point, a quantum-mechanical analogy is relevant %
as follows:\, when considering evolution of a system with %
time-dependent Hamiltonian \,$H(t)$\,, in general %
it would be absurd to confine consideration %
by such initial system's states (at \,$t=0$\,) %
 what are eigen-states of \,$H(0)$\,, or by %
such initial density matrix what commutes with \,$H(0)$\,.

If, nevertheless, in some application of the theory  %
jumps of FP seem unrealistic or ``bring %
unpleasantness'', then this indicates %
necessity to revise a system's model under use   %
but not FP's natural rights. %
Indeed, in any physically correct model characteristic %
temporal scales of system's reaction on external perturbations  %
must be determined by system itself,  %
but not by an outside ``censorship''.
So, to abandon FP's jumps would be as senseless  %
as to abandon the Heavyside step function or Green functions %
and other useful idealizations. %

Notice, besides, that in case of PP the same can be said %
about their time derivatives, \,$dx(t)/dt$\,, %
which also have rights to make jumps %
(that is \,$x(t)$\, may be continuously  %
piecewise linear time functions). %

\subsection{Comparison between JE and BKE}

According to p.2.4 and p.2.5, it remained only to consider %
the case CS-FP. Let us make this at \,$x(0)=0$\, when %
JE and BKE concern one and the same statistical ensemble, %
namely, defined by initial distribution \,$D_{eq}(\Gamma,0)$\,. %
Then we merely have to compare the random (fluctuating) %
quantities \,$W(t)$\, and \,$E(t)$\, in exponentials   %
of (\ref{ej}) and (\ref{ebk}). %
Forming their difference, from (\ref{w}) and (\ref{e}) or  %
(\ref{e0}) we have  %
\begin{eqnarray}
W(t)-E(t) = h(\Gamma,x(0)) - h(\Gamma(t),x(t))  %
= x(0)\cdot Q(0) - x(t)\cdot Q(t)\,\, \label{d}
\end{eqnarray}
with last expression corresponding to bilinear Hamiltonians. %
At that, taking into account that \,$x(0)=0$\,, one can %
rewrite BKE (\ref{ebk}) in confortable for the comparison form
\begin{eqnarray}
\langle\,\exp{\{\,-[W(t) + x(t)\cdot Q(t)]/T\}}\, %
\rangle_0\,=\,1\,\, \, \label{jbk}
\end{eqnarray}
where the random factor \,$-x(t)\cdot Q(t)$\, replaces %
the non-random \,$F(x(t))-F(0)$\,. %

Thus, JE and BKE deal with essentially different %
random quantities (functional of system's history)  %
and therefore mutually supplement one another. %
Differences between them disappear in such specific %
time moments only when \,$x(t)=x(0)=0$\,. But, obviously, %
it would be wrong to say about this that one of the two %
equalities ``reduces'' to another. %

Moreover, the statement that JE and BKE are %
``equivalent'' for ``cyclic processes'' \cite{jar} (see p.1.4) %
also is not  quite correct. This can be seen from following mental %
experiment. Let a parameter, - for instance, the pendulum's torque, - %
smoothly changes from \,$x(0)=0$\, to $x(t_0)=x_0\neq 0$\, %
and then rapidly, - during time \,$\delta t$\, %
much smaller than characteristic time scales of pendulum's %
motion, - returns to initial value: %
\,$x(t_0+\delta t)=0$\,. From (\ref{e0}) it is clear that %
the quantity \,$E$\, practically does not change during this %
return: \,$E(t_0+\delta t)-E(t_0)$ $\propto\delta t$\,. %
Simultaneously, according to (\ref{w}) and (\ref{d}), %
the quantity  \,$W$\, achieves the coincidence \,$W(t_0+\delta t)=$ %
$E(t_0+\delta t)$\, by means of jump %
\,$W(t_0+\delta t)-W(t_0)\approx $ $x_0\,Q(t_0)$\, %
practically independent on \,$\delta t$\,. %
Thus, one can say that in essence the \,$W$\,'s coincidence %
with \,$E$\, after ``cycling'' of the process  is nothing  %
but artifact having no relation to actual physical %
contents of these quantities.

It is useful also to consider two more special experiments. %
Let the torque grows from \,$x(0)=0$\, so slowly, - i.e. an  %
observation time is so long, - that the process %
can be considered as adiabatic. Then quantity \,$W(t)$\, %
in JE (\ref{ej}) is almost (asymptotically) free of fluctuations %
and merely reduces to the constant \,$\Delta F(t)$\,. %
Correspondingly, BKE (\ref{jbk}) turns to
\begin{eqnarray}
\exp{[-\Delta F(t)/T]}\,\,\langle\, %
\exp{[\,- x(t)\,Q(t)/T\,]}\, %
\rangle_{adiabatic}\, =\,1\, \,, \, \nonumber
\end{eqnarray}
which attests that fluctuations are quasi-equilibrium  in such %
process, i.e. \,$D(\Gamma;t)\approx$ $D_{eq}(\Gamma,x(t))$\,. %

Now, in opposite, let the torque sharply switches on  %
in the very beginning of observation and later stays constant:\, %
\,$x(t>0)=x =\,$const\,. Then \,$W(t>0)=$ $-x\,Q(0)=$\,const\,, %
where \,$Q(0)=Q(\Gamma(0))$\, obeys probability distribution %
\,$D_{eq}(\Gamma,0) $\,. Therefore JE degenerates into %
the bare \,$F(x)$\,'s definition, thus giving no information %
about actual changes in the system at \,$t>0$\,. %
At the same time, BKE gives %
\begin{eqnarray}
\langle\, %
\exp{[\,- x\,(Q(t)-Q(0))/T]}\, %
\rangle_{\, \uparrow \overrightarrow{\,\,\,\,x\,\,\,\,}} \, %
=\,1\, \, \, \label{jbk1}
\end{eqnarray}
Here, the arrows are symbolic image of assumed time dependence %
of the parameter. This is nontrivial equality, since its %
power expansion (or differentiation) in respect to \,$x$\, %
leads to ``Green-Kubo formulae'' and besides, - if elasticity %
of pendulum's wire (or ribbon) is not of Hooke's type,  %
or viscosity of the liquid is ``non-Newtonian'', and fluctuations are %
non-Gaussian, - to additional ``non-linear'' relations between %
fluctuations and dissipation. %
Thus one can see that general differences of BKE from JE %
are much more interesting than special cases of their %
``'equivalence''.

\subsection{On experimental testing of %
exact results of the theory}

If the authors of experiments described in \cite{tm} %
(or see \cite{lpp}) knowed about our old results, %
they would be able to test some of them, including %
BKE (\ref{ebk}), along with the JE (\ref{ej}) %
and Crooks equalities (we have to underline once  %
again that in case CS-FP old and new equalities are valid  %
simultaneously and independently one on another at %
any values of parameters). %
Or to test equality (\ref{jbk}) equivalent to BKE %
in application to the torsion pendulum. From (\ref{jbk}) %
it is clear that this even does not need in additional %
measurements. Especially simple for testing is particular %
case (\ref{jbk1}) of jump-like (step-like) torque %
switching on, which even does not require integration %
of measured data.

The same can be said about relations between probabilities %
of mutually time-reversed processes (see Section 3). %
In parallel with the ``Crooks equality'' \cite{lpp,cr1,cr2} %
one can test with torsion pendulum also relation %

\begin{eqnarray}
P(E;x)\,\exp{(-E/T)}\,=\, %
P(-E;\widetilde{x}) \,\,, \label{tm}
\end{eqnarray}
where \,$P(E;x)$\, is probability density distribution %
of the quantity \,$E=E(t)$\, under given %
parameter's trajectory \,$x=x(\tau)$\, ($0<\tau<t$),\,  %
\,$\widetilde{x}(\tau)= \epsilon x(t -\tau)$\, %
($\epsilon =\pm 1$), %
and initial micro-states on both sides  %
are subject (as in (\ref{ebk}), (\ref{jbk}) %
and (\ref{jbk1})) to the distribution %
\,$D_{eq}(\Gamma,0)$\,. %

This relation is direct consequence of above mentioned %
formula (7) from \cite{bk1} or other FDR for probability %
functionals (see p.3.2-4). %
In the example of pendulum \,$\epsilon =1$\, and, %
under designations of \cite{lpp}, \,$x=M$\,, \,$Q=\Theta $\,, %
\,$ E(t)=\int_0^t M(t^\prime)\, d\Theta(t^\prime)$\,, %
with \,$M$\, and \,$\Theta $\, being torque and rotation %
angle, respectively.  %

Let us consider two particular cases. In first of them, %
the torque switches on by jump and after that is constant %
like at end of previous paragraph. At that, equality %
(\ref{tm}) can be written similar to (\ref{jbk1}),
\begin{eqnarray}
P(E;\,\uparrow \overrightarrow{\,\,\,\,M\,\,\,\,})  %
\,\exp{(-E/T)}\,=\, %
P(-E;\,\uparrow \overrightarrow{\,\,\,\,M\,\,\,\,})\, %
\, \,, \label{tm1}
\end{eqnarray}
where \,$E$\, is merely \,$E(t)=(\Theta(t)-\Theta(0))\,M$\,. %
Notice that in this case both the forward and reversed  %
processes begin from zero torque, that is they are %
identical in statistical sense. Nevertheless, %
both processes are ``non-cyclic'' since \,$t$\, represents %
arbitrary time cut. If, however, a moment \,$t$\, %
is beforehand definitely stipulated, then one can make %
the processes formally ``cyclic'', replacing %
\,$\uparrow\overrightarrow{\,\,\,\,M\,\,\,\,}$\, %
by \,$\uparrow\overrightarrow{\,\,\,\,M\,\,\,\,}\downarrow $\,, %
all the more that such replacement does not influence %
on \,$E(t)$\, (see p.2.7). %

In the second case, let torque grows with time linearly. %
Then  in analogous symbolic notations formula (\ref{tm}) yields  %
\begin{eqnarray}
P(E;\,\nearrow\,)  %
\,\exp{(-E/T)}\,=\, %
P(-E;\,\uparrow \searrow\,)\, %
\, \,, \label{tm2}
\end{eqnarray}
where now forward and reversed processes are non-identical. %
At that, the reversed process begins by jump and looks like  %
``cyclic'', while the forward one is not such %
(though, again it can be made cyclic, by jump back  %
to zero, since \,$E(t)$\, is indifferent to this operation and %
thus \,$P(E;\nearrow\downarrow)=$ $P(E;\nearrow)$\,). %
Possibly, this is interesting variant for experimental  %
testing.

Anyway, we would like to notice that practical %
verification of principles and exact results of %
statistical mechanics is at the same time %
verification of validity of one or another model %
of a system under consideration. For example, %
 the mechanical external excitation of the torsion %
pendulum is performed with the help of electric current, %
which flows inside it, and magnetic field which %
pierces all the system. Therefore, interpretation %
of experiments in terms of the torque and rotation %
angle only means implicit assumption that all possible %
collateral effects of current and field are either weak %
enough or statistically independent on the observed %
mechanical motion of the pendulum.  %
More precisely, their non-considered addition  %
\,$E^\prime(t)$\, to the quantity \,$E(t)$\,, %
calculated from measurements of the torque and angle, %
is statistically non-correlated with \,$E(t)$\,, %
so that \,$\langle\exp{[-(E+E^\prime)/T]}\rangle$ %
$=\langle\exp{(-E/T)}\rangle$ %
$\langle\exp{(-E^\prime/T)}\rangle$\, %
and \,$\langle\exp{(-E^\prime/T)}\rangle=1$\,. %
Similarly, description of the experiments %
with RNA has attracted a Hamiltonian with PP, namely, %
position of the movable ``bead'' (see \cite{lpp}), %
although factually the system is governed through %
FP, such as voltage drop on the piezoelectric actuator. %
Thus again one assumes either weakness or statistical %
independence of collateral channels of  %
actuator-mediated external perturbation of the system. %

The ``independence'' in these examples %
is nothing but hypothesis that microscopic %
phase volume conserves separately in different %
(considered or ignored) channels of system's interaction %
with its surroundings. In reality, of course, in respect  %
to non-equilibrium processes this may be quite wrong %
assumption (see \cite{bkt}). %

\subsection{BKE for open systems,  %
or when JE is out of work}

After transition  from CS to OS, - for example, %
from torsion pendulum to rotator, - %
by the reasons expounded above in p.1.2, 2.1 and 2.4,   %
JE generally loses a sense. At that, the discussed %
in p.2.7 difference of the quantity \,$W(t)$\, (\ref{w}) %
in JE from quantity \,$E(t)$\, (\ref{e0}) in BKE  %
becomes dramatically aggravated.  %
Let us consider it in practically important situations %
to which Eqs.\ref{jbk1} and \ref{tm1} were addressed. %
The symbol \,$\uparrow$\, there now may denote either %
sharp switching on of FP (torque) \,$x=\,$const\,  %
or smooth switching on, during time interval \,$\delta t$\, %
smaller than rotator's period of revolution %
under given \,$x$\,

Now \,$E(t)\approx x\,\Delta Q(t)$\,, %
ith \,$\Delta Q(t)=Q(t)-Q(0)$\,, %
unboundedly grows with time,  approximately  %
linearly on average, if the external force %
gets balanced by viscous resistance of the liquid. %
At that \,$E(t)\approx x\,\Delta Q(t)$\, %
represents energy dissipated by the system in %
(quasi-) steady non-equilibrium state.  %
Analogous dissipative states arise, for example, %
when  \,$x$\, is electric potential applied to a conductor,  %
or force (electric field) acting on a particle %
(charge carrier) in unbounded (under thermodynamic %
limit) medium, while \,$\Delta Q(t)$ is transported charge  %
or particle's displacement. In all such situations %
BKE (\ref{ebk}) or (\ref{jbk1}) reveals definite %
rigid connections between generally non-linear %
dissipation (rheological properties of liquid, %
current-voltage characteristics of conductor,  %
particle's mobility, etc.) and statistical characteristics %
of fluctuations of \,$dQ(t)/dt$\, (angle velocity of rotator, %
electric current, particle's velocity, etc.), while  %
\,$E(t)$\, continuously accumulates new experimental %
information for these connections. %

In opposite, the quantity %
\,$W(t> \delta t)\approx$ $-x\,Q(0)=$\,const\,, %
like in p.2.7, remains constant in time, thus %
saying nothing about non-equilibrium processes in a system %
(moreover, even about \,$Q(0)$\,, since in OS  %
any value of \,$Q(0)$\, equally can be made reference %
point for \,$Q(t)$\,, by settling \,$Q(0)=0$\,). %
In order to extract from \,$W(t)$\, a portion %
of information about dissipated energy \,$E(t)$\,  %
at least for  a single time moment, one has, %
like in p.2.7, to retract the force \,$x$\, back to %
zero for a time, i.e. to arrange artificial %
``cyclic process'' and thus deteriorate the object %
(non-equilibrium steady state) under investigation. %
Moreover, according to p.2.7,  %
now because of the continuous growth \,$\Delta Q(t)$ $\propto t$\, %
with time the return to zero becomes more and more less %
correct operation, requiring more and more precise %
measurement of \,$x$\,, with error  %
\,$\sim T/\langle\Delta Q(t)\rangle =$ %
$x\,T/\langle E(t)\rangle$\,.
What is for continuous repetition of such operation, %
with hope to get from \,$W(t)$\, and JE information %
about the dissipative state, such an attempt would %
result merely in elimination of this state %
(that is genuine price of the ``equivalence''  %
of JE and BKE in ``cyclic processes''!).

\subsection{Evolution of non-equilibrium states %
and thermodynamical inequalities}

Let us consider CS which at \,$t=0$\, %
is in a non-equilibrium state.  We may try to model %
corresponding distribution \,$D(\Gamma;t=0)$\,  %
by an equivalent ``quasi-equilibrium'' one, %
\begin{eqnarray}
D_{qe}(\Gamma,X)=  %
\exp{\{\beta F^\prime(X)- [\beta H_0(\Gamma) + %
X\cdot Q(\Gamma)] \}} \,\,, \label{neq}
\end{eqnarray}
where \,$\beta =1/T$\,,\, \,$H_0(\Gamma )$\, is  %
eigen Hamiltonian of the system (in absence of external %
perturbations), \,$Q(\Gamma )$ is a suitable set %
of its individual or collective variables, \,$X$\, is %
conjugated set of parameters (``thermodynamic forces'') %
characterizing system's non-equilibrium, %
and  ``free energy'' \,$F^\prime(X)$\, %
is determined by the normalization condition.  %
The equivalence means that average values  %
of all the \,$Q(\Gamma )$\, over the factual %
and quasi-equilibrium distributions are coinciding.  %
This requirement determines all the \,$X$\,. %
The variables \,$Q(\Gamma )$\, may represent, for example, %
spatial inhomogeneities of densities of particles' number, %
mass, charge, momentum, energy, etc. %
Motivation of such quasi-equilibrium model  %
is that distribution (\ref{neq}) brings  %
to system's informational entropy maximum %
under given average values of \,$Q(\Gamma )$\,. %

In order to extend this model to other time moments, %
that is to system's evolution, it is natural %
to rely to formally exact statistical equalities. %
One of them, similar to BKE (\ref{ebk}), follows from identity %
(\ref{tj0}) at \,$D_1(\Gamma)=$ $D_2(\Gamma)=$  %
$D_{qe}(\Gamma,X)$\,. Namely, taking into account possible %
influence of external forces (fields) conjugated with some %
of the variables \,$Q(\Gamma)$\,, one obtains
\begin{eqnarray}
\langle\,\exp{[-\Delta S(t)]}\, %
\rangle\,=\,1\,\,, \label{nbk}
\end{eqnarray}
where angle brackets stand for averaging over %
quasi-equilibrium initial distribution, and  %
\begin{eqnarray}
\Delta S(t) = \beta E(t) + %
X\cdot [Q(t) - Q(0)] = \nonumber\\
= \int_0^t [\beta x(t^\prime) +X]\cdot %
\frac {dQ(t^\prime)}{dt^\prime}\, dt^\prime\,  \label{ndh}
\end{eqnarray}
This quantity usually can be treated as change, %
or increment, of entropy of a system in the course %
of its evolution. %

In \cite{bk8,bk3,bk4,pj} it was shown that statistical %
equalities and FDR associated with quasi-equilibrium %
ensembles of micro-states form a reliable base for %
nonlinear non-equilibrium thermodynamics. This is evident  %
already from such their simplest consequences as %
``thermodynamical inequalities''. %

It is well known that for any random quantity  %
\,$A$\, inequality\, %
$\langle \,\exp \,A\,\rangle \,\geq \, %
\exp \,\langle A\rangle $\, is true. %
Replacing here $\,A\,$ by $\,-E\,$, with quantity $\,E\,$ %
from (\ref{e}) or (\ref{e0}), and combining this inequality %
with equality (\ref{ebk}), it is not hard to conclude %
that\, $\langle E\rangle \,\geq 0$\,. %
Thus, if a system initially was equilibrium %
then it always on average (over statistical ensemble) %
takes from sources of external forces and absorbs a positive %
amount of energy.

Now, let a system is already initially non-equilibrium %
like above. Then equality (\ref{nbk}) implies  %
inequality $\,\langle \Delta S \rangle\,\geq 0$\,. %
It allows negative values of the average \,$\langle E\rangle $\,, %
that is now the system is able to produce (useful) work  %
against its surroundings, with  \,$-\langle E\rangle > 0 \,$\,. %
At that, the inequality $\,\langle \Delta S \rangle\,\geq 0\,$ %
together with (\ref{ndh}) establishes definite restriction on  %
value of this work, dependently on degree of system's %
initial non-equilibrium.

Next, let us go to statistical equalities including %
time-reversed processes, and demonstrate that  %
they also obey the presented in p.2.2 comparative %
characterization of ``old'' (our \cite{bk1,bk2,bk3}) and %
``new'' (Crooks \cite{cr1,cr2,cr3}) results. %


\section{Time reversibility of microscopic %
dynamics and generalized fluctuation-dissipation relations (FDR)}

\subsection{Time reversal}

In the classical mechanics, time reversibility of motion %
means that any phase trajectory of a system can be %
passed back in the time, if at some time moment %
\,$\theta $\, one inverts signs of all momenta (velocities). %
In case of non-autonomous system, one should also reverse %
time dependencies of external forces and conditions %
and besides invert signs of some of them (``odd'' parameters). %
In particular, sign of magnetic field and, under %
observations in a rotating frame (e.g. on the Earth's surface), %
sign of the Coriolis force. Parities, \,$\epsilon =\pm 1$\,, %
of corresponding Hamiltonian parameters in respect to time reversal %
are determined by requirement that %
\,$H(\overline{\Gamma},x)=$ $H(\Gamma,\epsilon x)$\,, %
where \,$\overline{\Gamma} \equiv $ $\{q, -p\}$\,. %
If it is satisfied, and in the ``forward'' time %
the dependence of current microscopic state  %
of a system, \,$\Gamma(t)$\,, on its initial state \,$\Gamma(0)$\, %
and on external conditions is expressed by functional %
\,$\Gamma(t)=$ $\mathcal{T}_t\{\Gamma(0);x(\tau)\}$\,, %
then in time-reversed view  %
\,$\Gamma(t)=$ $\overline{\mathcal{T}} %
_{\theta -t}\{\overline{\Gamma}(\theta); %
\epsilon x(\theta-\tau)\}$\,. Or, equivalently,  %
\,$\overline{\Gamma}(\theta -t)=$ %
$\mathcal{T}_t\{\overline{\Gamma}(\theta); %
\epsilon x(\theta-\tau)\}$\,, %
where \,$t$\, represents reversed time counted backward  %
from the ``turning point'' \,$\theta$\,. %

Correspondingly, observation of any variable \,$Q(\Gamma )$\,, %
which has definite parity, - %
\,$Q(\overline{\Gamma})=$ $\varepsilon Q(\Gamma)$\, %
($\varepsilon =\pm 1$), - %
instead of \,$Q(t)=Q(\Gamma(t))$\, on forward trajectory  %
gives \,$Q(\overline{\Gamma}(\theta -t))=$  %
$\varepsilon Q(\theta -t)$\, on the reversed trajectory. %


\subsection{Generalized FDR \label{p3_2}}

Let \,$\Phi\{\Gamma(\tau)\}$ be some functional of phase trajectory %
of a system, and let us consider its average value in  the %
ensemble of trajectories which is established by canonical %
initial distribution (\ref{eq}) with parameters \,$x=a$\,: %
\begin{eqnarray}
\langle \Phi\{\Gamma(\tau)\} %
\rangle_{a,\, x(\tau)} = %
\int \Phi\{\Gamma(\tau)\}\,  D_{eq}(\Gamma,a)\, %
 d\Gamma\,\, \nonumber 
\end{eqnarray}
Here the tag under angle brackets reminds %
about the initial distribution and external conditions %
at which the system evolves. Along with this average, consider %

\begin{eqnarray}
\int \Phi\{\Gamma(\tau)\}\, %
\exp{\left[\, \ln\, %
\frac {D_{eq}(\Gamma(\theta),b)} %
{D_{eq}(\Gamma,a)}\,\right]}\, %
D_{eq}(\Gamma,a)\, d\Gamma\, =\, %
\nonumber\\ \,= %
\int \Phi\{\Gamma(\tau)\}\, %
D_{eq}(\Gamma(\theta),b)\,
d\Gamma\,\, \label{f1}
\end{eqnarray}

\ The integrand here differs from integrand on the left %
in the identity (\ref{tj}) by additional multiplier %
\,$\Phi\{\Gamma(\tau)\}$ only. %
On right-hand side of (\ref{f1}), like in p.2.3, %
let us go from \,$\Gamma$\, to new integration variables,  %
now \,$\Gamma(\theta)$\,, then from them to %
\,$\overline{\Gamma}(\theta)$\,, and apply the Liouville %
theorem. After that, express the phase trajectory  %
via \,$\overline{\Gamma}(\theta)$\, while going  %
to reversed time, as it was described in previous paragraph.
Besides, redesignate the integration variable %
\,$\overline{\Gamma}(\theta)$\, by \,$\Gamma$\, %
and take into account that \,$D_{eq}(\overline{\Gamma},b) =$  %
$D_{eq}(\Gamma,\epsilon b)$\,. %
As the result, we obtain equality

\begin{eqnarray}
\left\langle \,\Phi\{\Gamma(\tau)\}\, %
\exp{\left[- \frac {H(\Gamma(\theta),b)- %
H(\Gamma,a)}T\right]}\, %
\right\rangle_{a,\, x(\tau)} \,=\, %
\nonumber\\ \,=\, %
\exp{\left[-\frac {F(b)-F(a)}T\right]}\, %
\, \langle \, \Phi\{\overline{\Gamma}(\theta -\tau)\} %
\,\rangle_{\epsilon b,\, \epsilon x(\theta -\tau)}  %
\,\, \label{sf}
\end{eqnarray}

\ At \,$\Phi\{\cdot\}=1$\, it reduces to the identity (\ref{tj}).

By choosing shape of the functional \,$\Phi\{\cdot\}$\, %
and parameters \,$a$\, and \,$b$\, in a proper fashion, %
it is easy to transform (\ref{sf}) into various FDR %
for characteristic and probability functionals  %
or statistical moments of system's variables.  %
In view of arbitrariness of \,$\Phi\{\cdot\}$\, %
equality (\ref{sf}) by itself is equipotent to visually %
similar relation for (density of) probability measure  %
in space of phase trajectories described with sufficient %
completeness (specification). The latter presumes that %
a set of variables, \,$V(t)=V(\Gamma(t))$\,, %
used in a ``coarse-grained'' description,  %
allows to express through themselves the difference %
\,$H(\Gamma(\theta),b)-H(\Gamma,a)$\, %
(though as for the rest may be arbitrary small %
or rough in comparison with  \,$\Gamma(t)$\,).
At that, from p.2.2 it follows that the parameters %
\,$a$\, and \,$b$\, must be chosen according to table %
\begin{eqnarray}
\begin{array}{ccc}
\texttt{OS} & \texttt{CS-FP} & \texttt{CS-PP} \\ %
\texttt{$a=b=0$} & \texttt{any\, $a$,$b$} &  %
\texttt{$a=x(0)$, $b=x(\theta)$}  %
\end{array} \label{vyb}
\end{eqnarray}
In the case CS-FP, any \,$a$\, and \,$b$\, are allowable, %
but nevertheless, as before in p.2.4-5 under the JE %
and BKE derivations, we will confine our consideration %
by the two particular variants from neighbor cells %
of this table.

So, when considering CS and taking \,$a$\, and \,$b$\, %
from third column of (\ref{vyb}), instead of (\ref{sf}) %
one can write
\begin{eqnarray}
\mathcal{P}[V;x]\,\,e^{-W(\theta)/T} \, \,=\, %
e^{-\Delta F(\theta)/T}\, %
\mathcal{P}[\widetilde{V}; \widetilde{x}]\, %
\,, \, \label{crf}
\end{eqnarray}
where \,$\mathcal{P}[V;x]$\, is (density of) probability %
distribution of possible observations (trajectories) %
of \,$V(\tau)$\, under given variations of parameters \,$x(\tau)$\, %
($0\leq \tau\leq \theta $),\, %
\,$\Delta F(\theta)= $ $F(x(\theta))-F(x(0))$,\, %
\,$\widetilde{V}(\tau)=$ $\varepsilon V(\theta -\tau)$\,,\, %
\,$\widetilde{x}(\tau)=$ $\epsilon x(\theta -\tau)$\,,\, %
and the quantity \,$W(\theta)$\, (change (\ref{w}) of %
system's full energy) is thought expressed via \,$V(\tau)$\,, %
that is variables \,$Q(t)$\, conjugated with \,$x(t)$\, %
are contained in the set \,$V(t)$\, or are functions of \,$V(t)$\,. %
This formula is equivalent of the %
``Crooks equality'' \cite{lpp,cr1,cr2,cr3,jar}. %

When considering OS, one has to take \,$a$\, and \,$b$\, %
from first column of (\ref{vyb}), which leads from (\ref{sf}) to  %
\begin{eqnarray}
P[V;x]\, \,e^{-E(\theta)/T} \,\,=\, %
P[\widetilde{V};\widetilde{x}]\,\, \,\label{p1}
\end{eqnarray}
Here \,$P[V;x]$\, has the same sense as above,\, %
and the quantity \,$E(\theta)$\, (change, (\ref{e}) or %
(\ref{e0}), of system's internal energy)  %
also is thought expressible in terms of \,$V(\tau)$\,. %
For this, it is more than sufficient if the Hamiltonian  %
of system-surroundings interaction, \,$-h(\Gamma,x)$\,,  %
can be written in the fom \,$-h(V(\Gamma),x)$\, %
(that is, for instance, if in the bilinear case %
(\ref{lh}) again \,$Q(t)$\, are either some of \,$V(t)$\, %
or some their functions). %
This formula is equivalent of formula (7) from \cite{bk1} %
and some formulae from \cite{bk2,bk3}. %

The probabilities in (\ref{crf}) and (\ref{p1}) are designated %
by different symbols because in large the equalities %
(\ref{crf}) and (\ref{p1}) relate to different statistical %
ensembles and different types of systems.  %
Among themselves they relate nearly like JE and BKE do (see p.2.7). %
Namely, (\ref{crf}) may be applied to OS, and (\ref{p1}) %
to case CS-PP, under condition \,$x(\theta)=x(0)=0$\, only, %
and then they formally coincide one with another. %
In the case CS-FP the two equalities do work simultaneously  %
and supplement one another, though being mutually connected  %
(as it will be shown in next paragraph). %

It is not hard to generalize equality (\ref{p1}) to  %
the quasi-equilibrium statistical ensemble described in %
p.2.10, with this purpose inserting to (\ref{f1}) %
distribution (\ref{neq}) in place of (\ref{eq}). %
As the result, in addition to (\ref{nbk}) one obtains
\begin{eqnarray}
P[V;x,X]\, e^{-\Delta S(\theta)}\,=\, %
P[\widetilde{V};\widetilde{x}, \widetilde{X}]\,\,,
\label{pt}
\end{eqnarray}
with quantity \,$\Delta S(t)$\,, defined  in (\ref{ndh}), %
and with \,$\widetilde{X}=\varepsilon X$\,. %

All the three relations (\ref{crf}), (\ref{p1}) and (\ref{pt}) %
can be unified into single relation like (\ref{mfds})  %
if in the first of them introduce \,$\Delta S =E/T$\,, %
in the second \,$\Delta S =$ $[W -\Delta F]/T$\,, in both of  %
them \,$\Pi_+ =$ $\{V;x\}$\,, %
\,$\Pi_- =$ $\{\widetilde{V};\widetilde{x}\}$\,, %
and similarly introduce \,$\Pi_\pm $\, for the third relation.
Notice, besides, that for ``mixed'' systems defined in p.2.2 %
(or systems having at once FP and PP) one has to (or may) write out  %
clear mix of equalities (\ref{p1}) and (\ref{crf}).%


\subsection{From old to new relations and back}

In the cases CS-FP the variables \,$V(t)$\, in ``new'' %
and ``old'' relations, - (\ref{crf}) and (\ref{p1}), - %
always can be identified.  Then differences between %
(\ref{crf}) and (\ref{p1}) reduce to difference of probability %
distributions of initial point \,$V_0=V(0)$\, of  %
trajectories \,$V(\tau)$\,. This means that
\begin{eqnarray}
\mathcal{P}[V;x]=\mathcal{P}[V|V_0;x]\, %
\mathcal{D}_{eq}(V_0,x(0))\,\,, \,\,\,\, %
P[V;x]=\mathcal{P}[V|V_0;x]\, %
\mathcal{D}_{eq}(V_0,0)\,\,, \label{div}
\end{eqnarray}
where \,$\mathcal{P}[V|V_0;x]$\, is common for %
(\ref{crf}) and (\ref{p1}) conditional probability %
distribution of trajectories \,$V(\tau)$\, %
under given their initial point, while  %
\begin{eqnarray}
\mathcal{D}_{eq}(v,a)\,= \int %
\delta(v-V(\Gamma))\, %
D_{eq}(\Gamma,a)\, d\Gamma \, \label{eq1}
\end{eqnarray}
are possible equilibrium distributions of the variables \,$V$\,. %
Indeed, in view of the completeness of these variables, %
from  (\ref{eq1}) and (\ref{eq}) it follows that %
\begin{eqnarray}
\mathcal{D}_{eq}(v,a) =  %
\mathcal{D}_{eq}(v,0)\,\exp\, %
\frac {F(a)-F(0)+h(v,a)}T\,\,, \nonumber 
\end{eqnarray}
where \,$h(V,a)$ is the interaction energy from  (\ref{fh}) %
expressed in terms of \,$V$\, (see comment after (\ref{p1})). %
Applying this equality, together with the decomposition %
(\ref{div}), to both sides of (\ref{crf}), after elementary %
manipulations subject to (\ref{d}) one arrives to (\ref{p1}). %
And, conversely, going from Eq.\ref{p1} by the same way %
but in opposite direction, one comes to equality (\ref{crf}).

Thus, from purely formal point of view, in the field of  %
CS-FP old and new relations are absolutely equivalent  %
at any time variations of parameters. %
But from the viewpoint of practical applications %
or testings they differ one from another %
as much essentially as JE and BKE do (see p.2.7-9). %

A literal practical realization of each of two %
variants of time reversal presumed in (\ref{crf}) and (\ref{p1})  %
requires preliminary preparation of system, %
that is its thermalization under correspomding %
Hamiltonian parameters from table (\ref{vyb}). %
In variant (\ref{p1}) transition from the preparation  %
of system to observation of its evolution may involve  %
jumps of Hamiltonian parameters in beginning  %
of observation (at leasr, in beginning of reversed %
process, similar to the example (\ref{tm2})).
We know from p.2.2 and p.2.6 that FP have %
all rights to make this. Such possibility %
is especially useful in case of OS, when the theory %
and its applications practically dispose of %
the ``old'' time reversal and FDR (\ref{p1}) only.


\subsection{Other FDR for probabilities  %
and fluctuation theorems}

If variables \,$V$\, do not form a complete set, %
then such one can be formed merely by adding to them %
the integral \,$W=W(\theta)$\, or \,$E=E(\theta)$\, %
as the whole. Then instead of (\ref{crf}) and (\ref{p1}) %
the equality (\ref{sf}) implies
\begin{eqnarray}
\mathcal{P}[V,W;x]\,\,e^{-W/T} \,=\, %
e^{-\Delta F/T}\, %
\mathcal{P}[\widetilde{V},-W; \widetilde{x}]\, %
\,, \, \label{crf1}\\
P[V,E;x]\,\,e^{-E/T} \,=\, %
P[\widetilde{V},-E; \widetilde{x}]\, %
\,, \, \label{p11}
\end{eqnarray}
where joint probability distributions of \,$V(\tau)$\, %
and \,$W=W(\theta)$\, or \,$E=E(\theta)$\, appear, %
and we took into account that values of \,$W$\, and  %
\,$E$\, in mutually time-reversed processes differ %
by signs only (it is easy to make sure of this with %
the help of ``time reversal rules'' from p.3.1).

Integration of (\ref{crf1}) and (\ref{p11}) %
over all (trajectories of) \,$V$\, yields relation %
\begin{eqnarray}
\mathcal{P}(W;x)\,e^{-W/T}\,= %
e^{-\Delta F/T} \mathcal{P}(-W; \widetilde{x})\, %
\,  \label{ft}
\end{eqnarray}
and above mentioned relation (\ref{tm}),  %
\,$P(E;x)$ $e^{-E/T} =$ $P(-E;\widetilde{x})$\,, %
 for marginal probability distributions of \,$W=W(\theta)$\, %
and \,$E=E(\theta)$\,, respectively. %
Then, dividing (\ref{crf1}) by (\ref{ft}), and  %
(\ref{tm}) by (\ref{p11}), one obtains relations  %
\begin{eqnarray}
\mathcal{P}[V|W;x]= %
\mathcal{P}[\widetilde{V}|-W; \widetilde{x}]\,\,,  %
\,\,\,\, P[V|E;x]= %
P[\widetilde{V}|-E; \widetilde{x}]\, %
\label{cp}
\end{eqnarray}
for conditional distributions of \,$V(\tau)$\,  %
under given values of \,$W=W(\theta)$\, %
or \,$E=E(\theta)$\, which here play role of additional %
external conditions of \,$V(\tau)$\,'s observations. %

Relations like (\ref{ft}) and (\ref{tm}) are very popular %
today and usually termed ``fluctuation theorems'' %
(FT) \cite{rmp,cht,pt}. Sometimes they are presented  %
as long forward step from ``old'' results. %
Although it is quite evident that equality (\ref{tm}) %
is simplest consequence of equality (\ref{p1}) %
produced by (mental) integration of (\ref{p1}) over %
all \,$V(\tau)$\,'s trajectories (\,$Q(\tau)$\,'s trajectories %
in \cite{bk1}) with fixed value of \,$E=E(\theta)$\,. %
Similarly, equality (\ref{pt}), or (\ref{mfds}), implies  %
FT for entropy increments:\, %
\begin{eqnarray}
P(\Delta S;x,X)\, \exp{(-\Delta S)}\,=\, %
 P(-\Delta S;\widetilde{x},\widetilde{X})\,\,
\nonumber 
\end{eqnarray}

We in our time wrote and used such relations %
in slightly different forms more comfortable %
in applications under our interest, taking into account  %
that really measured quantities usually are not  %
\,$E(t)$\, but \,$dQ(t)/dt$\, or \,$\Delta Q(t)$\, or %
others conjugated with external forces. %
Below in p.3.6 and p.3.8 it will be demonstrated %
how in 1979 in \cite{bk2} and later the FT (\ref{tm}) %
was applied by us to open systems (OS). %

Notice, besides, that in application to OS  %
in (quasi-) steady non-equilibrium states   %
the variant (\ref{tm1}) of FT (\ref{tm}) can  %
be rewritten as symmetry relation  %
\,$P(\sigma)$ $\exp{(-\sigma\theta)}=$ %
$P(-\sigma)$\, %
for time-averaged entropy production  %
\,$\sigma =$ $E(\theta)/T\theta$\,, %
and that such kind of relations may appear %
in non-Hamiltonian dynamic models of  %
dissipative processes \cite{ev,pt}. %
From viewpoint of the Hamiltonian statistical  %
mechanics, this  is advantage of such a model, %
though yet not proof of its legitimacy. %


\subsection{Marcovian FDR}

A set of variables \,$V(t)=V(\Gamma(t))$\, represents  %
a Marcovian random process if their current values  %
unambiguously and independently on their past determine  %
probabilities of their future values. %
The generalized FDR are formally compatible with %
assumptions about Marcovian behavior of one or another %
set of variables, although in rigorous sense it never %
takes place (unless if $V(\Gamma)$ coincides with $\Gamma$). %
This circumstance makes it possible to formulate %
general recipes for constructing such Marcovian %
``stochastic models'' which fully take into account  %
all FDR and thus automatically agree with both time %
reversibility of microdynamics and principles of  %
(statistical) thermodynamics of irreversible processes  %
\cite{bk1,bk2,bk3,bk4,bk5,bk6}. %
In respect to closed systems (CS) with constant parameters  %
this was made already by Stratonovich \cite{strm} %
(see also \cite{strb} and references therein and in  %
\cite{bk1,bk2,bk5}) basing on the principle of detailed  %
balance. In \cite{bk1} we showed that his results  can be  %
extended to non-constant (time-dependent) parameters. %

The corresponding ``old'' results wholly   %
contain the Crooks theory presented initially  %
as the Marcovian one \cite{cr1,cr2}.

In order to become convinced of this, let us recall   %
that a Markov process is completely determined by its  %
probabilities of transitions from \,$V_0=V(0)$\,  %
to \,$V_\theta =V(\theta)$\, during infinitesimally small  %
time \,$\theta\rightarrow 0$\,. %
Consider them, using for them notation  %
\,$\mathcal{P}(V_\theta |V_0;x)$\,. %
Notice that, firstly, due to the completeness of \,$V(t)$ %
(see p.\ref{p3_2}) at small \,$\theta$\, in (\ref{crf}) %
\,$W(\theta)\rightarrow $ %
$- dx(\theta)\cdot Q_0$\,, %
where \,$dx(\theta)=$ $x(\theta)-x(0)$ %
$\propto\theta$\,, and quantities \,$Q_0=Q(0)$\, %
(with \,$Q(t)$\,from (\ref{w})) are functions of %
\,$V(0)$\, or merely some of \,$V(0)$\,. %
Secondly, from (\ref{eq1}) and (\ref{eq}) it follows that   %
\,$\mathcal{D}_{eq}(V_\theta,x(\theta)) \rightarrow$  %
$\mathcal{D}_{eq}(V_\theta,x(0))$ %
$\exp{\{[\Delta F(\theta) + %
dx(\theta)\cdot Q_0]/T\}}$\,. %
Inserting these expressions, together with (\ref{div}), %
into (\ref{crf}), we see that the exponents in (\ref{crf}) %
cancel one another accurate to second order of  %
\,$\theta$\,, so that equality
\begin{eqnarray}
\mathcal{P}(V_\theta |V_0;x)\, %
\mathcal{D}_{eq}(V_0,x)\,=\, %
\mathcal{P}(\varepsilon V_0 | %
\varepsilon V_\theta; \epsilon x)\,
\mathcal{D}_{eq}(\varepsilon V_\theta, %
\epsilon x)\,\,  \label{db}
\end{eqnarray}
takes place, with \,$x=x(0)\approx x(\theta)$\,. %
It coincides with principle of detailed balance (PDB) for systems %
with constant parameters. This just means that the %
``Crooks equality'' is in essence time-nonlocal   %
formulation of the ``old'' Marcovian theory.

In its time-local formulation   %
\,$\mathcal{P}(V|V_0;x)\rightarrow $ %
$[1+\theta {\bf K}(V,\nabla,x)]$ $\delta(V-V_0)$\,, %
where \,$\nabla =$ $\partial/\partial V $\,, %
and transition probabilities from (\ref{db})  %
are replaced by ``kinetic operator'' \,${\bf K}(V,\nabla,x)$\,  %
and kinetic equation
\begin{eqnarray}
\dot{D}(V;t) = {\bf K}(V,\nabla,x(t))\, %
D(V;t)\,\,,  \label{ke}
\end{eqnarray}
in which \,$D(V;t)$ is current probability density distribution %
of the Marcovian variables. %
At that, role of equality (\ref{crf}) for CS, %
that is role of PDB (\ref{db}), is plaed  %
by operator-valued symmetry relation  %
\begin{eqnarray}
{\bf K}(V,\nabla,x)\, %
\mathcal{D}_{eq}(V,x)\, = \, %
\mathcal{D}_{eq}(V,x)\, %
{\bf K}^\dagger(\varepsilon V, %
\varepsilon \nabla,\epsilon x)\,\,, \label{cko}
\end{eqnarray}
which accumulates all consequences of microscopic %
reversibility and FDR. Here \,$\dagger$ is symbol of %
conjugation, or transposition, in the Sturm-Liouville sense. %
The operator \,${\bf K}$\, may be differential   %
(for example, in Fokker-Planck equations) as well as  %
 integral (for example, in Kolmogorov type equations). %
It should be noted that FDR in combination with the causality %
principle forbid any \,${\bf K}$\,'s dependence on time %
variations of parameters except an instant one, i.e. %
dependence on their current values only (but in no way on %
past values) \cite{bk1}\, %
\footnote{\, %
In general too the causality principle is very useful instrument  %
for analysis of consequences of generalized FDR \cite{bk1,bk2,bk3}. %
}\,. %
Stationary solutions of the kinetic equation (\ref{ke}) %
(at constant parameters) are the equilibrium distributions,  %
\,$\mathcal{D}_{eq}(V,x)$\, (\ref{eq1}). %

In 1978-1981 we suggested generalization of this Marcovian theory  %
to open systems (OS)  \cite{bk2,bk3,bk4,bk5,bk6}. %
Of course, it must be guided by relation (\ref{p1}) (or (\ref{pt})). %
In simplest Marcovian models of OS a set of variables  %
\,$V(t)=V(\Gamma(t))$\, is complete, thus allowing to express  %
through themselves the dissipated power (energy dissipated by %
system per unit time), i.e. the integrand in (\ref{e}) %
or (\ref{e0}) (for which even a single variable, e.g. %
\,$I(t)=$ $dQ(t)/dt$\,, may occur sufficient). %
Then instead of (\ref{cko}) one comes to essentially different %
operator-valued relation, %
\begin{eqnarray}
{\bf K}(V,\nabla,x)\, %
\mathcal{D}_{eq}(V,0)\, = \, %
\,\,\nonumber\\ =\, %
\mathcal{D}_{eq}(V,0)\, %
[{\bf K}^\dagger(\varepsilon V, %
\varepsilon \nabla,\epsilon x),+ %
{\bf N}(V,x)/T]\,\,, \label{oko}
\end{eqnarray}
where \,${\bf N}(V,x)\,$ represents the dissipated %
power (may be in the form \,${\bf N}(V,x)=$ %
$x\cdot I(V)$\,). %

Now stationary (when \,$x=$\,const) solution of kinetic  %
equation (\ref{ke}) is equilibrium (equals to  %
\,$\mathcal{D}_{eq}(V,0)$\,) at \,$x=0$\, only. %
If \,$x\neq 0$\, then stationary solution of (\ref{ke})  %
describes the above considered (see p.2.9 and p.3.4) steady %
non-equilibrium state with permanent entropy %
production \,\,$\langle\sigma\rangle =$ %
$\langle {\bf N}(V,x)/T\rangle$ $\neq 0$\,, %
when \,$E(t)/t \rightarrow$ %
$T\langle \sigma\rangle$\,, and fluctuations of \,$E(t)$\, %
and other quantities are characterized by violation of  %
balance of mutually time-reversed processes.

Of course, from (\ref{cko}) and (\ref{oko})  %
we can come back to (\ref{crf}) and (\ref{p1}), respectively.  %
And notice once again that for ``mixed'' systems,  %
i.e. possessing parameters of both closed and open type,  %
instead of equalities (\ref{cko})  or (\ref{oko}) %
it is necessary to write out their obvious hybrid. %



\subsection{FDR for transport processes}

Consider, for demonstration of one of possible applications %
of FDR (particularly, (\ref{p1}) and (\ref{tm1})), %
charge transport through a conductor under constant %
(after switching on at \,$t=0$\,) voltage drop \,$x$\,. %
Here dissipated energy \,$E(\theta)=x\Delta Q(\theta)$\,, %
with \,$\Delta Q(\theta) = $ $Q(\Gamma(\theta))-Q(\Gamma)$\, %
representing amount of charge transported through the  %
conductor during observation time. %

Let us combine exact FDR and a simple stochastic model of the system. %
The FDR will be delegated by relation (FT) (\ref{tm}) %
written in the form
\begin{eqnarray}
P(\Delta Q;x)\,\exp{(-x\Delta Q/T)}\,=\, %
P(-\Delta Q;x)\,\,, \label{ftq}
\end{eqnarray}
factually used in \cite{bk2}, where the probability %
distribution \,$P$\, now relates to the charge. %
What is for the model, assume that our conductor %
is a contact (like e.g. \,$p-n$\,-junction) and therefore %
charge is transported through it by discrete portions %
\,$\pm e$\, forming two opposite Poissonian random flows. %
This means that average value of the electric current\,, %
$\overline{I}(x) =\langle \Delta Q\rangle /\theta \,$\,, %
and spectral power density of the current noise, %
\,$S(x) =[\langle \Delta Q^2\rangle - %
\langle \Delta Q\rangle^2]/\theta \,$\,, %
are expressed by formulae %
\begin{eqnarray}
\overline{I}=e[n_+ -n_-] \,\,, \,\,\,\, %
S=e^2[n_+ +n_-]\,\,,\, \label{is}
\end{eqnarray}
in which \,$n_\pm =n_\pm (x)$\, are mean numbers %
of the elementary charge portions transferred %
per unit time in forward and backward directions. %

Clearly, FDR establish definite connection between %
\,$\, n_+ (x)$\, and \,$n_-(x)$\,. %
In \cite{bk2} it was extracted from relations for %
characteristic function of \,$\Delta Q$\, equivalent to (\ref{ftq}). %
Here, we merely can surmise that %
relation (\ref{ftq}) is valid not only in respect %
to \,$\Delta Q$\, as the whole but also in respect %
to elementary transfer events:\, %
\begin{eqnarray}
n_+(x)\,\exp{(-ex/T)}\,=\,n_-(x)\,\, \label{eft}
\end{eqnarray}
From here and from (\ref{is}) one obtains %
following relation between power of on-equilibrium noise %
and mean current (current-voltage characteristics): %
\begin{eqnarray}
S(x)=\,e\overline{I}(x)\, \coth{(ex/2T)}\,\, \label{ny}
\end{eqnarray}
At \,$e|x|\ll 2T$\, it reduces to the Nyquist formula %
for ``thermal noise'' while in the opposite case %
to the formula for ``shot noise''.

In such way FDR help to reveal %
universal connections between dissipative non-linearity %
of a transport process ($I(x)$), its noise characteristics %
($S(x)$), and type of its statistics. %
For Gaussian statistics, instead of (\ref{ny}) we  %
would obtain \,$S(x)=$ $2T\overline{I}(x)/x $\,. %
More complicated examples of this kind %
can be found in \cite{bk2,bk8,bk3,bk4,bkt,i1,i3,i2}. %

Notice that measurements of \,$\overline{I}(x)$, $S(x)$\, %
and higher-order cumulants of \,$\Delta Q$\, %
are in principle not worse way of experimental testing of %
of exact theoretical results, (\ref{ftq}), (\ref{tm}) %
and (\ref{ebk}), than one discussed in \cite{lpp} and in p.2.8.


\subsection{FDR for 1/f-noise}

Just considered stochastic model has principal defect: %
in it, the elementary random events have %
\,{\it a priori}\, (``in advance'') prescribed %
relative frequency (time-averaged number of events per unit time, %
or their ``probability per unit time''),\, %
$n_\pm (x)$\,, independent on concrete realization of experiment, %
i.e. on phase trajectory of a system. %
Although, as it was shown by Krylov many years ago %
\cite{kr}, statistical mechanics gives no grounds %
for such assumptions.  %

One can understand this statement already on intuitive level. %
Indeed, the mentioned assumption would be likely %
if the system remembered a number of past events %
and compensated its deviations from a ``norm'' %
by means of opposite deviation of number of later events. %
But this is impossible if the system forgets about events %
soon after they had happened. Then it does not %
distinguish between ``norm'' and ``deviation'' %
and therefore produces fluctuations in the number of events %
proportionally to its ``normal'' (average) value. %
This means that relative frequency of events %
(``probability per unit time'') undergoes %
low-frequency fluctuations with 1/f\,-type spectrum. %

For the first time similar reasonings were suggested and %
mathematically formulated in \cite{pjtf} and \cite{bk9,bk10} %
and later confirmed on the base of statistical mechanics %
in \cite{i1,i3,i2,dvr,p1,tmf,alt,last} and other works,  %
first of all in application to random walks %
(``Brownian motion'') of atomic-size particles. %

It should be underlined that the fluctuations (``1/f-noise'') %
of relative frequencies by their very nature %
do not violate existing (anyway prevalent) balance  %
or definite disbalance of mutually time-reversed %
events (processes). Therefore, - as it follows from generalized %
FDR \cite{bk9,bk10,i1,i2}, - various particular FDR like  %
(\ref{eft}) hold also for fluctuating relative frequencies %
and all derived ``kinetic'' quantities. For example, %
the Einstein relation \,$D=T\mu$\, between diffusivity and %
mobility of a walking particle, \,$D$\, and \,$\mu$\,, %
can be extended to their fluctuations \cite{i1} %
(as well as the Nyquist formula to fluctuations of conductance %
and fluctuations of ``instant'' spectral power density of %
``white'' electric noise \cite{bk9,i2}). At that, %
to substantiate such statements, the relation %
(\ref{ftq}) (formula (A4) from \cite{i1}) is quite sufficient. %

Due to these circumstances it is possible, - %
as suggested already in \cite{pjtf}, - %
to separate fast fluctuations (white noise) and %
low-frequency fluctuations (1/f-noise), making use of %
primitive phenomenological language but taking in mind %
its rigorous statistical-mechanical equivalent. %
Next, consider in such way statistics of random walk of %
probe (``marked'') gas particle, basing on results %
of \cite{i1,dvr,p1,tmf,alt}. %


\subsection{FDR and molecular Brownian motion}

Now, in the relation (\ref{ftq}) ((A4) from \cite{i1}) %
\,$\Delta Q$\, will denote displacement, or path, %
of ``Brownian particle'' (BP). %
Let \,$R$\, be projection of \,$\Delta Q$\, onto direction  %
of external force \,$x$\, applied to BP  (probe gas atom). %
In the widely known simplest stochastic model of Brownian motion,  %
the FDR (FT) (\ref{ftq}) is satisfied by the Gaussian distribution %
\begin{eqnarray}
P(R;x)=P_\mu(R;x)\equiv \frac %
{\exp{[-(R-\mu xt)^2/4T\mu t \,]}} %
{\sqrt{4\pi T\mu t}} \,\,\label{g} %
\end{eqnarray}
Here, it is assumed, of course, that the observation time %
\,$t\,$\, is much greater than BP's velocity relaxation time or %
mean free path time, \,$\tau \,$\,.

However, honest consideration of the exact BBGKY equations %
for infinite chain of many-particle distribution functions %
of a fluid shows that expression (\ref{g}) is incompatible  %
with absence of (contemporaneous) statistical correlations %
between the BP and gas atoms far distanced from it. %
A true expression (first obtained in \cite{p1} and then %
by different method in \cite{dvr,tmf}) can be represented %
by superposition of the Gaussian distributions with %
various values of BP's mobility:
\begin{eqnarray}
P(R;x)= \int_0^\infty P_\mu(R;x)\, %
U_t(\mu)\, d\mu\,\,, \,\, \label{ng}\\
U_t(\mu)=\frac {\overline{\mu}^2}{\mu^3}\, %
\exp{\left(-\frac {\overline{\mu}}\mu \right)}\, %
\Xi\left(\frac {T\mu}{v_0^2t}\right)\,\,,
\label{md} %
\end{eqnarray}
where $\Xi(\cdot)$ is a ``cut-off'' function which quickly %
vanishes at infinity and turns to unit at zero, $\Xi(0)=1$\,, %
and  \,$v_0$\, is characteristic thermal %
velocity of gas atoms (speed of sound). From here for %
variances of the BP's path and the dissipated energy %
\,$E=xR$\, we have  %
\begin{eqnarray}
\langle R,R\rangle = %
2T\overline{\mu}t + (\overline{\mu} xt)^2\, %
F\left(\ln{\frac t\tau }\right)\,\,, \label{r2}\\
\langle E,E\rangle = %
2T\langle E\rangle + \langle E\rangle^2\, %
F\left(\ln{\frac t\tau }\right)\,\,, \label{e2}
\end{eqnarray}
with \,$F(z) \approx \,z$\,. %
Here and below the angle brackets with \,$n$\, commas  %
inside denote joint  $(n+1)$-order cumulant of \,$n+1$\, %
random quantities separated by the commas %
(``Malakhov's cumulant brackets'' \cite{mal}). %
The second terms in (\ref{r2}) and (\ref{e2}) say about %
1/f\,-fluctuations of mobility  and dissipated power, respectively. %
At \,$x=0$\,, similar asymptotic characterizes %
fourth-order cumulant of \,$R$\,, thus reflecting %
identical fluctuations of BP's diffusivity %
\,$D=T\mu\,$ \cite{bk9,i2,pjtf}. %

The function \,$U_t(\mu)$\, (\ref{md}) is effective %
BP's mobility probability distribution. Its power-law %
long tail is generally typical for distributions %
accompanying 1/f-noise \cite{bk10,pjtf,alt}. At %
$\langle E\rangle =\overline{\mu} x^2t \gtrsim T$ %
this cubic tail manifests itself in the path distribution %
(\ref{ng}) on the right (if \,$x>0$\,)\,: %
\,$P(R;x)\approx \langle R\rangle^2/R^3$\, at  %
$\,R>\langle R\rangle$\,. Correspondingly, %
the similar tail appears in distribution of dissipated energy %
in (\ref{tm})\,: \,$P(E;x)\approx \langle E\rangle^2/E^3$\, %
at $\,E>\langle E\rangle$\,. %
Hence, probabilities of ``large deviations'' of the path %
and dissipated energy are highly maintained in %
comparison with that predicted by the Gaussian model (\ref{g}). %
Such distributions were many times observed %
in experiments with non-stationary photo-currents %
(charge injection currents) \cite{p1008}.

It is interesting that shortest way to these results %
runs from the FDR \cite{dvr} %
(though one can also find clear ways to them  from %
explicit virial expansions of non-equilibrium %
partition sums \cite{alt}). %
Let us choose in (\ref{sf}), at \,$a=b=0$\,,
\begin{eqnarray}
\Phi\{\Gamma(\tau)\} = %
\delta(Q(t)-R)\,\delta(Q) %
\prod_k [1+\phi(q_k)]\,\,, \nonumber
\end{eqnarray}
where \,$\phi(q)$\, is some function of atom's coordinates,  %
and the product is taken over all gas atoms %
(except the BP itself). %
Then, left side of (\ref{sf}) characterizes %
influence of initial spatial non-uniformity of gas %
onto BP's walk, while right-hand side %
describes statistical correlations between %
BP's path (during all the observation time) and current %
microstate of gas (in configurational space). %
Further, choosing the function \,$\phi(q)$\, properly, %
one can extract from (\ref{sf}) relation %
\begin{eqnarray}
\nu \, \frac {\partial P(R;x)}{\partial\nu}\, = %
P(R;x)\int [\,\nu(\rho|R;x)-\nu \,]\, d^3\rho\,\,, \label{vr}
\end{eqnarray}
where \,$\nu $\, is mean density %
(atoms' number concentration) of gas,\, %
and \,$\nu(\rho|R;x)\,$ is conditional average value of gas %
density at distance \,$\rho$\, from BP under %
given value of its path, %

From (\ref{vr}) it follows that

\begin{eqnarray}
\frac {\partial \ln\,P(R;x)}{\partial \ln\,\nu} \,\,>\, %
-\,\nu\Omega \,\,, \label{in}
\end{eqnarray}

\ where \,$\Omega$ is characteristic space volume %
to which the correlations of gas with BP do extend. %
On the other hand, in the Gaussian model (\ref{g}), %
subject to the known dependence $\,\mu\propto D\propto 1/\nu$\,, %
we have %
\begin{eqnarray}
\frac {\partial \ln\,P_\mu(R;x)} %
{\partial \ln\,\nu} \,= %
\frac 12 - \frac {\langle E\rangle}{4T}\, %
 \left[\left(\frac R{\langle R\rangle}\right)^2 %
-1\right]\,\,  \nonumber 
\end{eqnarray}
Comparing this expression with inequality (\ref{in}), %
we see that they in no way are compatible one %
with another, if the ``correlation volume'' %
$\,\Omega$\, is bounded above by a finite number. %
Hence, if the gas stays indiiferent to (forgets about) outcome, %
$\,R\,$, of BP's walk, then it is %
unable to suppress large values of \,$R$\, %
so categorically as the law (\ref{g}) does require. %

At the same time, the law (\ref{ng})-(\ref{md}) %
is well compatible with (\ref{in}), at\, \,$\Omega =2/\nu\,$\,. %
Of course, this (or other) value of \,$\Omega \,$\, %
can not be obtained from FDR themselves only, %
its calculation needs in the whole BBGKY hierarchy \cite{i1,p1} %
or equivalent means (see \cite{tmf,alt} %
and references in \cite{alt,last}). %


\subsection{Variance of dissipation fluctuations}

The previous paragraph gave example %
of large fluctuations of dissipation %
whose magnitude, according to (\ref{e2}), is of order of mean  %
dissipation value. Another such example was considered in \cite{i3}. %
If energy is dissipated through not one but many, \,$N\gg 1$\,, %
degrees of freedom, then variance of dissipation fluctuations, %
along with magnitude of power-law tail of their distribution, %
will be approximately \,$N$\, times smaller. %
Anyway, for systems with Hamiltonians of type (\ref{lh}) %
one can obtain \cite{i3} exact FDR
\begin{eqnarray}
\langle E\rangle =  %
\frac 1T \int_{1>2} x(1) %
\langle I(1),I(2)\rangle_{x(\tau)\eta(\tau-2)} %
\,\,x(2)\,d1d2\,\,, \label{e1}\\
\langle E,E\rangle = 2T\langle E\rangle \,+ %
\frac 2T \int_{1>2>3} x(1)\,x(2)\, \times\, \nonumber\\ %
\times\, \langle I(1),I(2),I(3)\rangle_{x(\tau)\eta(\tau-3)} %
\,\,x(3)\,d1d2d3\,\,\, \label{ee}
\end{eqnarray}
Here, the ciphers are replacement %
of literal time arguments (and their indices), %
\,$I(t)=dQ(t)/dt$\, are ``currents'' conjugated with %
external forces \,$x(t)$\,, and \,$\eta(t)$\,
is the Heavyside step function. Its presence   %
means that in second- and third-order cumulants in integrands %
the most early (right-hand) values of the ``currents'' %
\,$I(t)$\, (i.e. $\,I(2)\,$ and $\,I(3)\,$, respectively) belong %
to still undisturbed (equilibrium) system.  %
These equalities follow from general FDR for cumulants %
\cite{bk2,bk3} (importantly, as we already mentioned in %
p.2.2, \cite{bk3} contains two variants of such FDR,  %
which together cover both the types of systems and %
both the types of parameter). %

In many applications one can suppose that the third-order %
cumulant (or result of its integration) in (\ref{ee}) %
vanishes at \,$x=0$\,. Then, under weak perturbation  %
\begin{eqnarray}
\langle E,E\rangle = 2T\langle E\rangle \,+ %
\frac 2T \int_{\stackrel{1>4}{2>3,4}}\, %
x(1)\,x(2)\, \times\nonumber\\ %
\times\, \langle I(1),G(2,3),I(4)\rangle_{0}\,\,  %
x(3)\,x(4)\,d1d2d3d4\,\,,\, \label{eee}
\end{eqnarray}
where\, \,$G(2,3)=[\delta I(2)/\delta x(3)]_{x=0}$\,\, %
is dynamical (that is introduced at the level %
of microscopic dynamics) differential linear %
response of the currents to the forces \cite{i3}, %
the first term \,$\propto x^2$\,, and the second \,$\propto x^4$\,. %
But the second term does not reduce to mere small correction,  %
if it grows with observation time faster then linearly. %
For instance, like in (\ref{e2}), approximately  %
$\propto t^2$\,. Then the integral in (\ref{eee}) %
describes contribution from equilibrium 1/f\,-fluctuations %
of kinetic coefficients of the system. %
As it is visible from (\ref{eee}) %
(and demonstrated in \cite{i3} and \cite{last}), %
they in natural way are related to fluctuations %
of the differential response \,$\delta I(2)/\delta x(3)$\,, %
characterizing exponential instability of %
of system's phase trajectories in respect to  %
their small perturbations \cite{arn,kr,lili}.

\section{Conclusion}

We have presented to readers our view of the   %
generalized fluctuation-dissipation relations (FDR),  %
or theorems, for the first time introduced by us %
in 1977, in their comparison with analogous results  %
what appeared in 1997 and later. %
All they manifest phase space volume conservation  %
under microscopic dynamic motion and its time symmetry %
(reversibility), both substantially determining %
statistical and dissipative properties %
of thermodynamically non-equilibrium physical %
systems\, %
\footnote{\, %
In the context of FDR, the word ``thermodynamic'' and %
its derivatives concern mainly statistical %
ensembles but not sizes of systems under interest, %
so that the generalized FDR are equally valid  %
for both large systems (even with infinite number %
of degrees of freedom) and small ones (even with %
single degree of freedom) \cite{bk1}. %
}\,. %

The comparison gave us sufficient reasons to say, - %
in contrast to the misunderstandings %
observed in related literature (see Introduction), - %
that ``new'' results have not introduced a principal %
novelty or greater generality, in essence appearing  %
alternative formulations of ``old'' results. %
Our approach suggested in our time  %
(and reflected in this paper) has more general %
character, helping to notice and take into account %
qualitative peculiarities of different types of systems, %
first of all on the level of their %
Hamiltonians and statistical ensembles, %
and then their stochastic models. %
In the framework of our approach %
one easy can see inter-connections of new %
and old results, possibilities to choose %
most adequate form of FDR for concrete %
application, and to derive new variants %
of FDR not considered before.

On the whole, the generalized FDR %
bring all necessary tools %
for construction of thermodynamically correct %
models of real non-equilibrium processes and systems. %
Regardless of degree of complexity or roughness %
of a model, observance of FDR at its level %
ensures its qualitative agreement with rigorous %
statements of statistical mechanics %
(and sometimes even leads closely %
to quantitative agreement, as was demonstrated, %
in particular, by examples in last %
paragraphs of the present paper). %
Hardly this useful potential of FGR %
will be exhausted some day. %


\appendix

\section*{Appendix}

The following Appendix is absent in Russian original %
of this present submitted to Physics-Uspekhi, %
but may be useful ``pedagogical'' supplement to it.

\subsection{From particular to general FDR}

Let us consider a linear dissipative conductor %
(``resistor'') with conductance \,$G$\, at temperature \,$T$\, %
under external voltage drop \,$x(t)$\, causing %
current \,$I(t)$\,. Neglecting detail frequency dispersion of  %
conductivity, but taking into account the Nyquist formula %
along with the Ohm's law, one can write
\begin{eqnarray}
\langle I(t)\rangle_{x(\tau)} = G\,x(t-0)\,\,, \nonumber \\ %
\langle I(t), I(t^\prime)\rangle_{x(\tau)} = %
2TG \delta(t-t^\prime)\,=\, TG\, [\delta(t-t^\prime -0) + %
\delta(t-t^\prime +0)] \,\, \label{sm}
\end{eqnarray}
for ensemble-average value of the current and correlation %
function (second-order cumulant function) of its thermal %
fluctuations, respectively. %
Here, zero in the argument \,$t-0$\, at top row  %
means (infinitely) small positive number and %
reminds that because of the causality principle  %
\,$I(t)$\,'s response to \,$x(t)$\, possesses at least  %
a little time delay. Correspondingly, delta-function  %
in bottom row consists of retarded and advance parts. %
Neglecting also non-Gaussianity of the fluctuations, %
one comes to the current's characteristic %
functional as follows,
\begin{eqnarray}
\langle \exp{[\int iu(t)\,I(t)\, %
dt]}\rangle_{x(\tau)} \,=\, 
\exp{\{\int iu(t)\,TG\,[\,iu(t-0) + %
x(t-0)/T]\, dt\}} \, \label{cf0}
\end{eqnarray}
Here \,$iu(t)$\, is arbitrary probe function (generally %
complex-valued). %

This characteristic functional (CF) reoresents very %
particular stochastic model of an open system (OS), %
but undoubtedly physically correct model %
in those sense that the nature really may %
produce random processes arbitrary close to pure Gaussiqn %
ones, in accordance with the ``central limit theorem''. %
Therefore, observations made from Eq.\ref{cf0} can %
occur to be of general significance. %
First of such observations is rather evident. Namely, %
it is easy to see that under special choice %
\,$iu(t)=-x(t)/T$\, Eq.\ref{cf0} reduces to equality
\begin{eqnarray}
\langle \exp{[-\int I(t)\, x(t)\, %
dt/T]}\rangle_{x(\tau)} \,=\, 1 \,\, \label{ie}
\end{eqnarray}
Notice, besides, that this equality contains no signs %
of above assumptions, i.e. linearity and time locality %
of conductance, and Gaussianity and ``white noise'' %
character of current fluctuations. %
This observation prompts that Eq.\ref{ie} %
has very general meaning and must be valid regardless of %
actual voltage-current response and statistics of current noise. %
Thus, we in fact have come to the BKE (\ref{ebk}). %

If so, then we may expect that other properties of %
particular  CF (\ref{cf0}) also are extendable to general case  %
if they can be written irrespective to its specificity. %
Such a property appears when one makes in (\ref{cf0}) change %
of the probe variable (function) %
\,$iu(t) \rightarrow \, iu(t)-x(t)/T$\,. %
After it the exponent in (\ref{cf0}) transforms as %
\[
\begin{array}{c}
\int [\,iu(t)-x(t)/T]\,TG\,iu(t-0)\, dt\,  = \\
= \int (-iu(-t^\prime))\,TG\,[-iu(-t^\prime -0) %
+x(-t^\prime -0)/T]\, dt^\prime\,\,,
\end{array}
\]
where also change of the integration variable, %
\,$t =-(t^\prime -0)$\,, that is time reversal, %
is made. It is necessary in order to restore  %
thr correct cause-and-consequence disposition of %
\,$iu(-t^\prime)$\, (``consequence'') relative to %
\,$x(-t^\prime)$\, (``cause''). Comparison between %
this expression and (\ref{cf0}) yields equality %
\begin{eqnarray}
\langle \,\exp{\{\int [\,iu(t)-x(t)/T]\,I(t)\, %
dt\,\}}\,\rangle_{x(\tau)} \,=  \nonumber \\
=\,\langle\, \exp{\{\int (-i\epsilon u(-t))\, %
I(t)\, dt\,\}}\,\rangle_{\epsilon x(-\tau)}\,= \, %
\label{ire} \\ =\, %
=\,\langle \,\exp{\{\int iu(t)\, %
(-\epsilon I(-t))\, %
dt\,\}}\,\rangle_{\epsilon x(-\tau)}\,\,, \nonumber
\end{eqnarray}
where \,$\epsilon =1$\,.
In view of arbitrariness of the probe function, %
it is clear that this relation is equivalent %
to Eq.\ref{p1}, with \,$V(t)=I(t)$\,. %

Thus, starting from the Nyquist formula in context %
of most primitive stochastic model, and then %
reformulating it in most abstract terms of characteristic  %
functionals (CF), it is possible to reveal very general FDR %
((\ref{ie}) and (\ref{ire})) applicable %
to much more complicated stochastic models (with %
arbitrary non-linearity, frequency dispersion, non-Gaussianity, etc.). %
In fact, just these observations stimulated us %
thirty six years ago to recognize more fundamental %
statistical-mechanical derivation of Eq.\ref{ebk},  %
Eq.\ref{p1} and other generalized FDR \cite{bk1,bk2}.

Notice that under special choice \,$u(t)=\xi x(t)$\, %
Eq.\ref{ire} simplifies to
\begin{eqnarray}
\langle \,e^{\,(i\xi -\beta)\,E}\, %
\rangle_{x(\tau)} \,=\, %
\langle \,e^{-i\xi\,E}\, %
\rangle_{\epsilon x(-\tau)}\,\,, \,  \label{tmc}
\end{eqnarray}
with \,$E=\int x(t)\,I(t)\,dt$\, and \,$\beta=1/T$\,. %
Clearly, this is equivalent of the FT (\ref{tm}) %
which results from Eq.\ref{tmc} after its Fourier transform %
over \,$\xi$\,.


\subsection{Quasi-equilibrium correlators for OS}

The particular expression (\ref{cf0}) foresees also general %
structure of CF of ``currents'' \,$I(t)=$ $dQ(t)/dt$\, %
in OS found in \cite{bk2}:
\begin{eqnarray}
\langle \,\exp{[\int iu(t)\,I(t)\, dt]}\,\rangle_{x(\tau)} %
\,=\,  \label{cfo}\\ =\, %
\exp{\,\int_{1>2} iu(1)\,T\,G_{1,2} %
\{iu(\tau);\,x(\tau)\}\,[\,iu(2)+x(2)/T]} \, %
d1\,d2\,\, \nonumber
\end{eqnarray}
Here ciphers in place of indexed letters are used %
like in p.3.9, and \,$G{1,2}\{\cdot\}$\, is functional %
of fragments of trajectories \,$iu(\tau)$\, and \,$x(\tau)$\, %
with \,$2<\tau<1$\,. It satisfies time symmetry relation %
\begin{eqnarray}
G_{1,2}\{\,iu(\tau)-x(\tau)/T;\,x(\tau)\}\, %
=\, \nonumber\\ \,=\, %
G_{\theta-2,\,\theta-1} %
\{\,-i\epsilon u(\theta-\tau);\, %
\epsilon x(\theta-\tau)\}\,\,, \label{sro}
\end{eqnarray}
with formally arbitrary time shift \,$\theta$\,, %
e.g. \,$\theta =0$\,. It is easy to show \cite{bk2} that %
such CF's structure follows from the causality %
principle as combined with FDR (\ref{ire}) (or (\ref{p1})). %
One can see also that because of the condition \,$2<\tau<1$\, %
contributions from term %
\,$iu(1)$ $TG_{1,2}\{iu(\tau);x(\tau)\}$ $iu(2)$\, %
to (second- and higher-order) currents' cumulants %
\,$\langle I(t_1),\dots ,I(t_n)\rangle$\, depend %
on \,$x(\tau)$\, with \,$\tau\geq $ %
$\min{(t_1,\dots ,t_n)}$\, only. %
This means that
\begin{eqnarray}
TG_{1,2}\{\,iu(\tau);\,x(\tau)\}\, =\, %
\langle I(1),I(2)\rangle_{x(\tau)\,\eta(\tau -2)}\,+ %
\nonumber\\ + %
\int_{1>3>2} \langle I(1),I(3),I(2)\rangle_ %
{x(\tau)\,\eta(\tau -2)}\,\,iu(3)\, d3\,+\, %
\label{cfoe}\\ + %
\int_{1>3>4>2} \langle I(1),I(3),I(4),I(2)\rangle_ %
{x(\tau)\,\eta(\tau -2)}\,\,iu(3)\,iu(4)\, d3\,d4\, %
\,+\, \dots\,\,, \nonumber
\end{eqnarray}
with \,$\eta(t)$\, being the Heavyside function. %
Such the correlators (cumulants) can be named %
``quasi-equilibrium'' since most early current value there, %
\,$I(2)$\,, represents still equilibrium system, %
as if the force \,$x(t)$\, (the voltage or some other) %
was zero at \,$t <t_2$\,. They were in use above in %
p.3.9 (see also \cite{i3}). %

Insertion of (\ref{cfoe}) into (\ref{cfo}) yields %
simple but seemingly non-trivial expansion of %
non-equilibrium cumulants over the quasi-equilibrium ones: %

\begin{eqnarray}
\langle\, I(t)\,\rangle_ %
{x(\tau)}\,=\, %
\frac 1T \int^{t}_{-\infty}  %
\langle\, I(t)\,,\,I(t^\prime)\,\rangle_  %
{x(\tau)\,\eta(\tau -t^\prime)}\,\, %
x(t^\prime)\, dt^\prime\,\, %
\, \label{ceo1}
\end{eqnarray}
and, at \,$n\geq 2$\,,

\begin{eqnarray}
\langle\, I(t_1),\dots ,I(t_n)\,\rangle_ %
{x(\tau)}\,=\, \langle\, I(t_1),\dots ,I(t_n)\,\rangle_
{x(\tau)\,\eta(\tau -t_{min})}\,+ %
\nonumber\\ + %
\,\frac 1T \int^{t_{min}}_{-\infty}  %
\langle\, I(t_1),\dots ,I(t_n),I(t)\,\rangle_  %
{x(\tau)\,\eta(\tau -t)}\,\, x(t)\, dt\,\,, %
\, \label{ceo}
\end{eqnarray}

\ where \,$t_{min}=\min{(t_1,\dots ,t_n)}$\,. %
Then symmetry properties of all the correlators are determined %
by relation (\ref{sro}). %


\subsection{From time-local Marcovian to nonlocal formulation %
of generalized FDR}

Let us recall that if an equation like (\ref{ke}) from p.3.5 %
is ``kinetic'', that is its solution,
\begin{eqnarray}
D(V;t)=\, \overleftarrow{\exp}\, %
\{\int_0^t {\bf K}(V,\nabla,x(\tau))\, %
d\tau\,\}\, D(V;0)\,\,,  \label{kes}
\end{eqnarray}
represents evolution of normalized probability distribution %
of (Marcovian) variables \,$V(t)$\,, %
then solution to modified equation
\begin{eqnarray}
\dot{D}^\prime (V;t) = %
[\,\Psi(t,V) + {\bf K}(V,\nabla,x(t))\,]\, %
D^\prime(V;t)\,\,,  \label{gke}
\end{eqnarray}
with same initial condition, gives CF of %
variables \,$\Psi(t,V(t))$\,. Namely,
\begin{eqnarray}
D^\prime(V;t)=\, \overleftarrow{\exp}\, %
\{\int_0^t \! [\Psi(\tau,V) + {\bf K}(V,\nabla,x(\tau))]\, %
d\tau\,\}\, D(V;0)\,=\, %
\nonumber\\ =\, %
D(V;t)\, \langle\, \exp\, \int_0^t \! %
\Psi(\tau,V(\tau))\, d\tau \,\rangle_ %
{x(\tau)}^{V(t)=V}\,\,\,, \, \label{gkes}
\end{eqnarray}
where the angle brackets with additional tag above them %
represent conditional average, under condition that %
\,$V(t)=V$\,. Correspondingly,
\begin{eqnarray}
\int D^\prime(V;t)\, dV\,=\, \nonumber\\ =\, %
\int \! \overleftarrow{\exp}\, %
\{\int_0^t [\,\Psi(\tau,V) + {\bf K}(V,\nabla,x(\tau))]\, %
d\tau\,\}\, D(V;0)\, dV\,=\, %
\nonumber\\ =\, %
\langle\, \exp\, \int_0^t %
\Psi(\tau,V(\tau))\, d\tau \,\rangle_ %
{x(\tau)}\,\, \label{mcf}
\end{eqnarray}
is usual unconditional CF of \,$\Psi(t,V(t))$\,. %
At that, formally \,$\Psi(t,V)$\, may be arbitrary function %
for which the average (\ref{gkes}) is finite. %

Taking this facts in mind, first, assume that the evolution (kinetic) %
operator \,${\bf K}$\, possesses the symmetry property %
(\ref{cko}) characterizing closed systems (CS). %
Second, choose initial condition to Eq.\ref{gke} to be %
the \,$V$\,'s equilibrium distribution %
with \,$x=x(0)$\,, i.e.
\[
\begin{array}{c}
D^\prime(V;0)=D(V;0) =\mathcal{D}_{eq}(V,x(0))\,\,,
\end{array}
\]
with distribution \,$\mathcal{D}_{eq}$\, defined %
by Eq.\ref{eq1}. Third, write solution %
to Eq.\ref{gke} in the form
\[
\begin{array}{c}
D^\prime(V;t)=\, \mathcal{D}_{eq}(V,x(t))\,U(V;t)\,\,
\end{array}
\]
with \,$U(V;0)=1$\,, simultaneously applying relation %
(\ref{cko}) (time-local formulation of generalized 
FDR). Then Eq.\ref{gke} turns to
\begin{eqnarray}
\dot{U}(V;t) = %
[\Psi(t,V) - d(\ln{ %
\mathcal{D}_{eq}(V,x(t))})/dt\, %
 + {\bf K}^\dagger(\varepsilon V, %
\varepsilon \nabla,\epsilon x(t))\,]\, %
U(V;t)\,\,  \label{gke_}
\end{eqnarray}
Fourth, notice that action of left side of the %
operator-valued relation (\ref{cko}) onto unit produces %
zero (since \,$\mathcal{D}_{eq}(V,x)$\, is stationary %
solution of Eq.\ref{ke} at \,$x=$\,const\,), %
therefore always
\begin{eqnarray}
{\bf K}^\dagger(\varepsilon V, %
\varepsilon \nabla,\epsilon x)\,1\, %
=\,0 \,\, \label{cons}
\end{eqnarray}
(in essence this identity expresses conservation of total %
probability and probability distribution normalization %
during evolution). Consequently, if in Eq.\ref{gke_} we choose %
\,$\Psi(t,V)=$ $d\ln{\mathcal{D}_{eq}  %
(V,x(t))}/dt\,$\, then its %
solution is \,$U(V;t)=1$\,. %
From viewpoint of Eqs.\ref{gkes} and \ref{mcf} this means, evidently, %
that \,$D^\prime(V;t)=$ $\mathcal{D}_{eq}(V,x(t))$\,, %
hence,
\begin{eqnarray}
\mathcal{D}_{eq}(V,x(t))\,=\, %
D(V;t)\, \langle\, \exp\, \int_0^t %
\frac {dx(\tau)}{d\tau} \cdot %
\frac {\partial \ln\,\mathcal{D}_{eq} %
(V(\tau),x(\tau))}{\partial x(\tau)}\, %
d\tau \,\,\rangle_ %
{x(\tau)}^{V(t)=V}\,\,, \label{gkes_}\\
1\,=\,\langle\, \exp\, \int_0^t %
\frac {dx(\tau)}{d\tau} \cdot %
\frac {\partial \ln\,\mathcal{D}_{eq} %
(V(\tau),x(\tau))}{\partial x(\tau)}\, %
d\tau \,\,\rangle_{x(\tau)}\,\, \label{mcf_}
\end{eqnarray}
Fifth, notice that from Eq.\ref{eq1} and \,$V$\,'s %
completeness (see formula after Eq.\ref{eq1}) it follows that
\begin{eqnarray}
\frac {\partial \ln\,\mathcal{D}_{eq}(V,x)}{\partial x}\,=\, %
\frac 1T\, [\,\frac {dF(x)}{dx}\, %
+\,Q(V,x)\,]\,\,, \, \label{dld}
\end{eqnarray}
with \,$Q(V,x)=\int Q(\Gamma,x)\, %
\delta(V-V(\Gamma))\,D_{eq}(\Gamma,x)\,d\Gamma\, /$  %
$\mathcal{D}_{eq}(V,x)\,$ %
$=\,\partial h(V,x)/\partial x$\,, %
being the  variables conjugated with \,$x$\, and %
expressed in terms of \,$V$\,. %
At last, inserting (\ref{dld}) to (\ref{mcf_}), %
we come to the Jarzynski equality (JE) (\ref{ej}). %

By the way, inserting (\ref{dld}) to (\ref{gkes_}), %
we obtain interesting relation
\begin{eqnarray}
\mathcal{D}_{eq}(V,x(t))\,=\, %
D(V;t)\,\, e^{\Delta F(t)/T}\, %
\langle\,e^{-W(t)/T}\,\rangle_ %
{x(\tau)}^{V(t)=V}\,\,, \label{nd}
\end{eqnarray}
with designations from p.2.5 %
(\,$\Delta F(t)=$ $F(x(t))-F(x(0))$\,, %
\,$W(t)=$ $\int_0^t Q(\tau)\cdot dx(\tau)$\,, %
\,$Q(t)=$ $Q(V(t),x(t))$\,). %
It connects the non-equilibrium probability %
distribution \,$D(V;t)$\, to the quasi-equilibrium %
one and conditional average value %
of the exponential \,$\exp{(-W(t)/T)}$\,. %
Such kind of FDR for the first time was %
considered in \cite{bk2,bk3}. %

In order to obtain more general non-local FDR, %
let us take
\begin{eqnarray}
\Psi(t,V)\,=\, iu(t)\cdot V\,+\, %
\frac {d\ln{\mathcal{D}_{eq}(V,x(t))}}{dt} \, \,, \label{ch_}
\end{eqnarray}
with \,$iu(t)$\, being arbitrary probe functions. %
Then solution to Eq.\ref{gke_} is
\begin{eqnarray}
U(V;t)=\, \overleftarrow{\exp}\, %
\{\int_0^t [\,iu(\tau)\cdot V\, %
 + {\bf K}^\dagger(\varepsilon V, %
\varepsilon \nabla,\epsilon x(\tau))\,]\, %
d\tau\,\}\,1 \,\, \nonumber
\end{eqnarray}
After its substitution to top row %
of Eq.\ref{mcf}, transposition of the  %
operator exponential when integrating over \,$V\,$, %
transition from resulting anti-chronological exponential %
to chronological one, and then change of the integration  %
variables to \,$\varepsilon V\,$, we have
\begin{eqnarray}
\int D^\prime(V;t)\, dV\,= %
\int \mathcal{D}_{eq}(V,x(t))\,U(V;t)\, %
dV\,=\, \nonumber\\ =\, %
\int \overleftarrow{\exp}\, %
\{\int_0^t [\,i\varepsilon u(t-\tau)\cdot V\, %
 + {\bf K}(V,\nabla,\epsilon x(t-\tau))]\,  %
d\tau\,\}\, \,\mathcal{D}_{eq}(V,\epsilon x(t))\, %
dV\,\, \nonumber  
\end{eqnarray}
(here equality \,$\mathcal{D}_{eq} %
(\varepsilon V,x)=$ $\mathcal{D}_{eq}(V,\epsilon x)$\, %
is also taken into account). Obviously, from viewpoint of %
Eq.\ref{mcf}, this is nothing but  \,$V(t)$\,'s CF %
for time-inverted processes. %
At that, in bottom row of Eq.\ref{mcf} we have %
again CF of \,$\Psi(t,V)$\, but now chosen %
as in (\ref{ch_}). Thus, wholly Eq.\ref{mcf} %
now yields

\begin{eqnarray}
\langle\, \exp\, \int_0^t %
i\varepsilon u(t-\tau)\cdot V(\tau)\, %
d\tau \,\,\rangle_{\epsilon x(t-\tau)}\, %
\,=\, \label{mcf1}\\ \,=\, %
\langle\, \exp\, \int_0^t %
[\,iu(\tau)\cdot V(\tau)\,+\, %
\frac {dx(\tau)}{d\tau} \cdot %
\frac {\partial \ln\,\mathcal{D}_{eq} %
(V(\tau),x(\tau))}{\partial x(\tau)}\,] %
\,d\tau \,\,\rangle_{x(\tau)}\,\, \nonumber %
\end{eqnarray}

\ This FDR can be written also as

\begin{eqnarray}
\langle\, \exp\, \int_0^t %
iu(\tau)\cdot \varepsilon V(t-\tau)\, %
d\tau \,\,\rangle_{\epsilon x(t-\tau)}\, %
\,=\, \label{mcf1_}\\ \,=\, %
\langle\, \exp\, \int_0^t %
[\,iu(\tau)\cdot V(\tau)\,+\, %
\frac {dx(\tau)}{d\tau} \cdot %
\frac {\partial \ln\,\mathcal{D}_{eq} %
(V(\tau),x(\tau))}{\partial x(\tau)}\,] %
\,d\tau \,\,\rangle_{x(\tau)}\,\, \nonumber %
\end{eqnarray}

\ Performing here (mentally) functional Fourier transform in respect %
to \,$u(\tau)$\,, we can replace this FDR for CF %
by equivalent FDR for probability functionals,

\begin{eqnarray}
\mathcal{P}[\widetilde{V};\,\widetilde{x}]\,=\, %
\mathcal{P}[V; x]\, %
\langle\, \exp\, \int_0^t %
\frac {\partial \ln\,\mathcal{D}_{eq} %
(V(\tau),x(\tau))}{\partial x(\tau)}  \cdot  %
dx(\tau) \,\,\rangle_{x(\tau)}^{V(\tau)}\,
\,, \,  \label{mpf}
\end{eqnarray}

\ where \,$\widetilde{x}(\tau)=\epsilon x(t-\tau)$\,,  %
\,$\widetilde{V}(\tau)=\varepsilon V(t-\tau)$\,, %
and angle brackets denote conditional average %
under given trajectories \,$x(\tau)$\, and %
\,$V(\tau)$\, (at \,$0\leq\tau\leq t$,).

If \,$V$\, is a complete set of variables, in the sense %
of p.3.2, then \,$\partial\,\ln\,\mathcal{D}_{eq} %
(V,x)/\partial x$\, again reduces to (\ref{dld}), %
therefore expression in the angle brackets in  %
Eq.\ref{mpf} becomes conditionally non-random,
\begin{eqnarray}
\langle\, \exp\, \int_0^t %
\frac {\partial \ln\,\mathcal{D}_{eq} %
(V(\tau),x(\tau))}{\partial x(\tau)}  \cdot  %
dx(\tau) \,\,\rangle_{x(\tau)}^{V(\tau)}\,\rightarrow\, %
\exp{\,\frac {\Delta F(t)-W(t)}T}\,\,, \,\nonumber
\end{eqnarray}
and thus Eq.\ref{mpf} coincides with the ``Crooks %
equality'' (\ref{crf}) (with \,$t$\, in place of %
\,$\theta$\,). This is the case, in particular, %
if \,$V$\,'s are the same as full set of micro-variables \,$\Gamma$\,. %
Then ``kinetic operator'' \,${\bf K}$\, can be %
identified with the Liouville evolution operator\, %
\footnote{\, %
Just such approach was exploited in \cite{bk3}. %
At that, generally speaking, the \,$x$\,'s are introduced %
as parameters of the evolution operator (not of %
a Hamiltonian since it may not appear in theory at all). %

Formally, of course, Marcovian (stochastic) theory %
(dynamics) is more general than Hamiltonian %
(deterministic) one. But principally the latter %
is more adequate to the nature since it by itself produces  %
all possible randomness, including 1/f-noise  %
(i.e. ``randomness of degree of randomness and rate of %
dissipation'') what may be lost or ``killed'' in Marcovian models. %

Notice, besides, that approach based on evolution operator %
naturally allows its generalization to quantum (Hamiltonian)  %
case (to be considered separately elsewhere).
}\,. %
If, however, \,$V$\,'s are not complete, then %
Eqs.\ref{mcf1}-\ref{mcf1_} or \ref{mpf} give some generalization of Eq.\ref{crf}. %

Derivation (and similar generalizations) of Eqs.\ref{p1} %
and \ref{tm} for OS, basing on the symmetry relation %
(\ref{oko}) instead of (\ref{cko}) and formulae %
(\ref{gkes})-(\ref{mcf}), is even more simple task %
than just considered one.


\subsection{Quasi-equilibrium correlators for CS}

Let in previous paragraph \,$Q\in V$\, %
or \,$Q=Q(V,x)$\,, that is \,$V$\, is certainly complete set, %
and \,$B\in V$\, or \,$B=B(V,x)$\, are some additional to %
\,$Q$\, variables with definite parities. %
Then derivation {\it \,ad exemplum\,} that of Eq.\ref{mcf1} %
implies %

\begin{eqnarray}
\langle\, \exp\, \int_0^t %
[\,i\varepsilon w(t-\tau)\cdot B(\tau)\, %
+\, i\epsilon u(t-\tau)\cdot Q(\tau)\,]\, %
d\tau \,\,\rangle_{\epsilon x(t-\tau)}\, %
\,=\, \label{mcf2}\\ \,=\, %
e^{\Delta F(t)/T}\, %
\langle\, \exp\, \int_0^t %
\{\,iw(\tau)\cdot B(\tau)\,+\, %
[\,iu(\tau)+\dot{x}(\tau)/T]\cdot Q(\tau)\,\}\,  %
\,d\tau \,\,\rangle_{x(\tau)}\,\, \nonumber %
\end{eqnarray}

\ Here additional probe functions \,$w(t)$\, are introduced, %
and formula (\ref{dld}) is taken into account. %
Notice that \,$\int_0^t \dot{x}\cdot %
Q\,d\tau\,=$ $-W(t)$\,.

Then let us pay attention to %
that Eq.\ref{mcf2} remains valid %
if we add to integrands in its left and right-hand %
sides terms \,$-i\epsilon u(t-\tau)\cdot$ %
$Q_{0}(\epsilon x(t-\tau))$\, and %
\,$-i u(\tau)\cdot$ $Q_{0}(x(\tau))$\,, respectively, %
with any \,$Q_{0}(x)$\, satisfying %
\,$Q_{0}(\epsilon x)=$ $\epsilon Q_{0}(x)$\,. %
Let us choose it to be \,$Q_0(x)=Q_{eq}(x)$\, with
\begin{eqnarray}
Q_{eq}(x)\,\equiv \,\int Q(V,x)\, %
\mathcal{D}_{eq}(V,x)\, dV\,=\, %
-\, \frac {dF(x)}{dx}\,\,, \, \label{q0}
\end{eqnarray}
so that \,$Q_{eq}(\epsilon x)=\epsilon Q_{eq}(x)$\, %
by definition. Clearly, \,$Q_{eq}(x)$\, %
are equilibrium mean values of variables \,$Q(t)$\, %
at constant parameters, and last equality %
in Eq.\ref{q0} is direct consequence of %
Eq.\ref{dld} %
\footnote{\, %
In fact, in the framework of Marcovian theory %
in itself (without appeals to Hamiltonian dynamics) %
Eq.\ref{dld} serves as definition of the variables %
\,$Q\,$ conjugated with external parameters \,$x$\,. %
}\,. %
Besides, add to the same integrands terms %
\,$-i\varepsilon w(t-\tau)\cdot$ %
$B_{eq}(\epsilon x(t-\tau))$\, and %
\,$-i w(\tau)\cdot$ $B_{eq}(x(\tau))$\,, respectively, %
with
\[
\begin{array}{c}
B_{eq}(x)\,\equiv \,\int B(V,x)\, %
\mathcal{D}_{eq}(V,x)\, dV\,=\, %
\varepsilon B_{eq}(\epsilon x)\,
\end{array}
\]
Next, notice that nothing prevents to move the %
initial time moment from zero to arbitrary far past time, %
while the final moment arbitrarily far to the future. %
After these manipulations Eq.\ref{mcf2} takes form %
\begin{eqnarray}
\langle\, \exp\, \int %
[\,i\varepsilon w(\theta -\tau)\cdot \hat{B}(\tau)\, %
+\, i\epsilon u(\theta -\tau)\cdot \hat{Q}(\tau)\,]\, %
d\tau \,\,\rangle_{\epsilon x(\theta-\tau)}\, %
\,=\, \label{mcf3}\\ \,=\, %
\langle\, \exp\, \int %
\{\,iw(\tau)\cdot \hat{B}(\tau)\,+\, %
[\,iu(\tau)+\dot{x}(\tau)/T]\cdot \hat{Q}(\tau)\,\}\,  %
\,d\tau \,\,\rangle_{x(\tau)}\,\,, \nonumber %
\end{eqnarray}
with\, \,$\hat{Q}(t)\equiv Q(t)-Q_{eq}(x(t))$\,\, %
(\,$\hat{B}(t)\equiv B(t)-B_{eq}(x(t))$\,) being %
\,$Q(t)$\,'s ($B(t)$\,'s) deviations from their quasi-equilibrium %
values \,$Q_{eq}(x(t))$\, (\,$B_{eq}(x(t))$\,), and %
\,$\theta\,$ arbitrary constant. At \,$w(\tau)=$ %
$u(\tau)=0$\, this relation reduces to   %
\,$\langle \exp\int \hat{Q}(\tau)\cdot %
dx(\tau)\,\rangle $ $=1$\,, i.e. to the JE (\ref{ej}). %

FDR (\ref{mcf3}) implies definite restrictions on structure %
of \,$Q(t)$\,'s CF, similar to ones considered above %
in Appendix.1-2. To see them, we have to repeat the reasonings %
expounded in \cite{bk2}. In particular, they lead to %
expression like Eq.\ref{cfo},
\begin{eqnarray}
\langle \,\exp{[\int iu(t)\,\hat{Q}(t)\, dt\,]}\,\rangle_{x(\tau)} %
\,=\, %
\nonumber\\ =\, %
\exp{\,\int_{1>2} iu(1)\,S_{1,2} %
\{\,iu(\tau);\,x(\tau)\}\,[iu(2)-\dot{x}(2)/T]} \, %
d1\,d2\,\, \label{cfc}
\end{eqnarray}
Here for simplicity we took \,$w(\tau)=0$\,. %
The functional \,$S_{1,2}\{\cdot\}$\, again depends on  %
\,$iu(\tau)$\,, \,$x(\tau)$\, with \,$2<\tau<1$\, only %
and satisfies symmetry relation resembling (\ref{sro}), %
\begin{eqnarray}
S_{1,2}\{\,iu(\tau)+\dot{x}(\tau)/T;\,x(\tau)\}\, =\, %
S_{\theta-2,\,\theta-1} %
\{i\epsilon u(\theta-\tau);\, %
\epsilon x(\theta-\tau)\}\,\, \label{src}
\end{eqnarray}
To extend these expressions, Eqs.\ref{cfc} and %
\ref{src}, to  to non-zero \,$w(\tau)$\,, %
it is sufficient to replace arrays (vectors) %
\,$\hat{Q}$\, and \,$x$\, by %
\,$\{\hat{Q},\hat{B}\}$\, and \,$\{x,0\}$\,. %
In other words, one may merely treat \,$B(t)$'s\, %
as a part of \,$Q(t)$'s\,, namely, such part whose %
conjugated parameters are identically zeros. %

The CF's structure in Eq.\ref{cfc} again clearly %
reflects the causality principle, saying that %
mean value of \,$\hat{Q}(t)$\, can differ from zero %
only if parameters were changing somewhen before, i.e. %
\,$\dot{x}(\tau)\neq 0$\, at some \,$\tau <t$\,. %
Similarly to Eq.\ref{cfoe}, the functional  %
\,$S_{1,2}\{iu(\tau);x(\tau)\}$\, is composed of %
``quasi-equilibrium'' correlators (cumulants), %
that is presuming no external perturbations %
before beginning of observation (most early time argument  %
of a correlator). But now, in CS,  perturbations are %
characterized by rates of parameter's changes, %
\,$\dot{x}(t)=$ $dx(t)/dt$\,, instead of their deviations from %
zero in case of OS. Correspondingly,
\begin{eqnarray}
S_{1,2}\{\,iu(\tau);\,x(\tau)\}\, =\, %
\langle \hat{Q}(1),\,\hat{Q}(2)\rangle_ %
{[\,x(\tau)-x(2)]\,\eta(\tau -2)+ x(2)}\,+ %
\nonumber\\ + %
\int_{1>3>2} \langle \hat{Q}(1),\hat{Q}(3),\hat{Q}(2)\rangle_ %
{[\,x(\tau)-x(2)]\,\eta(\tau -2)+ x(2)}\,\,iu(3)\, d3\,+\, %
\label{cfce}\\ + %
\int_{1>3>4>2} \langle \hat{Q}(1), %
\hat{Q}(3),\hat{Q}(4),\hat{Q}(2)\rangle_ %
{[\,x(\tau)-x(2)]\,\eta(\tau -2)+ x(2)}\,\, iu(3)\,iu(4)\, d3\,d4\, %
\,+\, \dots\,\, \nonumber
\end{eqnarray}
Such cumulants,\,
\[
\langle \hat{Q}(t_1),\dots %
,\hat{Q}(t_n)\rangle_{[\,x(\tau)-x(t_{min})]\, %
\eta(\tau -t_{min})+ x(t_{min})}\,\,\,\,\,  %
\,\,\,\,(\,\,
t_{min}\,=\,\min{(t_1,\dots ,t_n)}\,\,)\, , %
\]
were considered already in \cite{i3}. %
It is necessary to underline that at any \,$n\geq 2$\,
\[
\langle \,\hat{Q}(t_1)\,,\,\dots\, ,\,\hat{Q}(t_n)\,\rangle\,=\, %
\langle\, Q(t_1)\,,\,\dots\, ,\,Q(t_n)\,\rangle \,\,,
\]
since non-random constituents of quantities subject to second- %
and higher-order cumulants, e.g. \,$Q_{eq}(x(t_j))$\,,  %
do not contribute to them. %
Taking this in mind, inserting (\ref{cfce}) to (\ref{cfc}) %
and expanding the result into power series over %
\,$iu(\tau)$\,, we come to representation of \,$n$\,-order %
(\,$n\geq 2$\,) non-equilibrium %
cumulants via quasi-equilibrium ones,
\begin{eqnarray}
\langle\, Q(t_1),\dots ,Q(t_n)\,\rangle_ %
{x(\tau)}\,=\, \langle\, Q(t_1),\dots ,Q(t_n)\,\rangle_
{[\,x(\tau)-x(t_{min})]\, %
\eta(\tau -t_{min})+ x(t_{min})}\,-\,  %
\nonumber\\ - %
\,\frac 1T \int^{t_{min}}_{-\infty}  %
\langle\, Q(t_1),\dots ,Q(t_n),Q(t)\,\rangle_  %
{[\,x(\tau)-x(t)]\, %
\eta(\tau -t)+ x(t)}\,\, %
\frac {dx(t)}{dt}\, dt\,\, \, \label{cec}
\end{eqnarray}
with\, \,$t_{min}=\min{(t_1,\dots ,t_n)}$\,. %
At \,$n=1$\, Eqs.\ref{cfc} and \ref{cfce} yield
\begin{eqnarray}
\langle\, \hat{Q}(t)\,\rangle_{x(\tau)}\,=\, %
-\,\frac 1T \int^{t}_{-\infty}  %
\langle\, Q(t),\,Q(t^\prime)\,\rangle_  %
{[\,x(\tau)-x(t^\prime)]\,\eta(\tau -t^\prime)+%
x(t^\prime)}\,\, \frac {dx(t^\prime)}{dt^\prime}\, %
dt^\prime\,\, \, \nonumber
\end{eqnarray}
or, equivalently,
\begin{eqnarray}
\langle\, Q(t)\,\rangle_{x(\tau)}\,=\, %
Q_{eq}(x(t))\,-\, \frac 1T \int^{t}_{-\infty}  %
\langle\, Q(t),\,Q(t^\prime)\,\rangle_  %
{[\,x(\tau)-x(t^\prime)]\,\eta(\tau -t^\prime)+%
x(t^\prime)}\,\, \frac {dx(t^\prime)}{dt^\prime}\, %
dt^\prime\,\, \, \label{cec1}
\end{eqnarray}
This formula gives evident exact decomposition of %
non-equilibrium mean values into quasi-equilibrium part %
and correction to it due to past variations of parameters. %
Thus, the correction always can be exactly expressed %
through pair (second-order) quasi-equilibrium cumulant %
(a kind of nonlinear extension of FDT). %

Getting back the variables \,$B(t)$\,, - as was %
explained above, - we come from Eqs.\ref{cec}-\ref{cec1} %
to formulae (2.25)-(2.26) from \cite{bk3} mentioned %
in p.2.2 and p.3.9 (with those only difference that in Eq.2.26 %
in \cite{bk3} the case of bilinear Hamiltonians was displayed).  %


\subsection{On some omitted reservations}

When writing about very wide field of FDT it is  %
impossible to mention all its potentially important aspects. %
Here we would like to point out briefly %
some of that omitted in the body of this paper. %

1. Of course, in reality one can meet systems what are %
neither strictly closed nor strictly open. %
For instance, if in the case of rotator (see p.2.1) %
its ``eigen'' Hamiltonian from Eq.\ref{lh} %
looks in fact like
\[
\begin{array}{c}
H_0(\Gamma^\prime,Q) = %
H_0^\prime(\Gamma^\prime) + %
h_0\,\cos\,Q\,\,, \,
\end{array}
\]
then the system is formally open at any %
\,$x \neq 0$\,, but %
at \,$|x|<|h_0|$\, it may stay in long-living %
meta-stable states and behave mainly as a closed %
system, though with non-Hooke's elasticity.

2. We qualified systems as ``closed'' or ``open'' %
dependently on their reaction to constant ``force %
parameters'' (FP) (see p.2.2). At the same time, of course, %
practically all systems are open in respect to %
constantly oscillating perturbations %
represented by oscillations of either FP %
or ``position parameters''(PP).

Of course, all general FDR are equally applicable %
to arbitrary (quasi-) periodic perturbations, %
at least (in classical variant under present %
consideration) with frequencies \,$\ll \hbar/T$\,. %
Notice that FDR from p.2.9 were applied in %
\cite{i3} just to periodically varying parameters.

3. Although systems with constant PP are definitely closed, %
continuous monotonous changes of PP may settle them %
into (quasi-) steady non-equilibrium state (SNS). %
For example, if rotation angle \,$Q$\, of the rotator %
is turned into external PP (so that force it applies %
to the fluid becomes conjugated internal variable), %
then its linear change \,$Q(t)= \omega\,t$\, clearly %
drives the system in an SNS. At that %
\,$\omega(t)=$ $dQ(t)/dt$\, has all rights to make jumps. %

4. If dear reader have recognized some other %
probably significant aspects of the subject, %
this does not mean that they are unknown to us.  %


\end{document}